# Global 3D hydrodynamic modeling of in-transit Lyα absorption of GJ436b

M. L. Khodachenko[1], I. F. Shaikhislamov[2], H. Lammer[1], A. G. Berezutsky[2], I. B. Miroshnichenko[2], M. S. Rumenskikh[2], K. G. Kislyakova[3]

1) Space Research Institute, Austrian Acad. Sci., Graz, Austria
2) Institute of Laser Physics SB RAS, Novosibirsk, Russia
3) Dep. of Astrophysics, University of Vienna

E-mail address: ildars@ngs.ru

**Abstract:** Using a global 3D, fully self-consistent, multi-fluid hydrodynamic model, we simulate the escaping upper atmosphere of the warm Neptune GJ436b, driven by the stellar XUV radiation impact and gravitational forces and interacting with the stellar wind. Under the typical parameters of XUV flux and stellar wind plasma expected for GJ436, we calculate in-transit absorption in Lyα and find that it is produced mostly by Energetic Neutral Atoms outside of the planetary Roche lobe, due to the resonant thermal line broadening. At the same time, the influence of radiation pressure has been shown to be insignificant. The modelled absorption is in good agreement with the observations and reveals such features as strong asymmetry between blue and red wings of the absorbed Lyα line profile, deep transit depth in the high velocity blue part of the line reaching more than 70%, and the timing of early ingress. On the other hand, the model produces significantly deeper and longer egress than in observations, indicating that there might be other processes and factors, still not accounted, that affect the interaction between the planetary escaping material and the stellar wind. At the same time, it is possible that the observational data, collected in different measurement campaigns, are affected by strong variations of the stellar wind parameters between the visits, and therefore, they cannot be reproduced altogether with the single set of model parameters.

## 1. Introduction

A series of observations with the *Hubble Space Telescope / Space Telescope Imaging Spectrograph* (HST/STIS) (*Kulow et al. 2014, Ehrenreich et al 2015, Lavie et al. 2017*) revealed that the warm Neptune GJ436b has a very deep transit in Lyα line with up to 60% of the stellar flux in this line being absorbed. Moreover, this strong absorption takes place mostly in the blue wing of the line in the range of Doppler shifted velocities of [-120; -40] km/s. The good signal-to-noise ratio allowed to clearly reveal for the first time in VUV observations of close-orbit hot/warm exoplanets, such details of the transit light-curve as early ingress (*Ehrenreich et al 2015*) and extended egress (*Lavie et al. 2017*). The quality of obtained data enables quantitative testing of existing theoretical concepts and numerical models proposed hitherto for GJ 436b and other similar close-orbit exoplanets.

The basic physical concept, behind the measured absorption in the Lyα line greatly exceeding the optical transit depth and duration, is related with the expanding hydrogen-dominated upper atmospheres of close-orbiting exoplanets. According to the energy limited estimates (*Lammer et al. 2003*), the ionizing radiation of a host star leads at orbital distances <0.2 a.u. to the intensive thermal escape and mass loss of hydrogen dominated upper atmospheres of hot gas giants. Since the first detection of an excess absorption in the Lyα of the hot-jupiter HD209458b at the level of 10% (*Vidal-Majar at al. 2003*), this concept has been rapidly developed, with continuously increasing complexity of numerical models. The first generation of 1D aeronomy codes (*Yelle*

*2004, García Muñoz 2007, Koskinen et al. 2007*) clarified the basic physics of the escaping upper atmosphere in the form of planetary wind (further PW), which includes the XUV heating, hydrogen plasma photo-chemistry, radiation cooling, gravitational and thermal pressure forces. They helped to explain some of the in-transit spectral observations by the presence of an expanded partially ionized upper atmospheres, which fill the Roche lobes of hot giant exoplanets, such as HD209458b and HD189733b (*Ben-Jaffel 2007, Ben-Jaffel & Sona Hosseini 2010, Koskinen et al. 2007, 2010*). These expanding atmospheres were shown to be sufficiently dense to produce the absorption in Lyα due to natural line broadening mechanism.

However, the detection of absorption in the resonant lines of heavy elements such as OI, CII, and SiIII (*Vidal-Madjar et al. 2004, Linsky et al. 2010*), has shown that the absorbing material of planetary origin far beyond the Roche lobe has to be considered as well (*Ben-Jaffel & Sona Hosseini 2010, Shaikhislamov et al. 2018a*). The presence of a huge hydrogen corona also is a prerequisite for the explanation of strong in-transit Lyα absorption of GJ436b. By this, there is another crucial factor, besides of the Roche lobe effect, which has to be properly taken into account in the modeling of large-scale plasma dynamics around the close-orbit exoplanets – the stellar wind (further SW) plasma. Self-consistent description of the escaping multi-component PW flow, accelerated by the pressure gradients and stellar gravity, and its interaction with the SW required an upgrade of the first generation of 1D models to 2D and 3D ones. The corresponding effort has been undertaken in *Shaikhislamov et al. (2016, 2018a,b)* and *Khodachenko et al. (2017)*.

Outside the Roche lobe, under the conditions of escaping planetary upper atmospheric material (hydrogen and/or other elements), the dominating absorption mechanism is the thermal line broadening, provided by the resonant particles moving at the corresponding matching speeds. Thus, for e.g., Lyα absorption at Doppler-shifted velocities, up to hundred km/s, the presence of hydrogen energetic neutral atoms (ENAs) is required, as the atoms of PW are two slow and cold and they cannot contribute this process. Two mechanisms for the generation of ENAs were proposed: 1) acceleration by the stellar radiation pressure, which in case of hydrogen is provided by the stellar Lyα flux (*Vidal-Majar et al. 2003*), and 2) charge exchange between fast stellar protons and planetary atoms (*Holmstrom et al. 2008*).

Taking the appropriate values for the parameters of PW, passing the Roche lobe, the generation and further dynamics of ENAs can be simulated with the kinetic Monte-Carlo models, to interpret the transit observations. Such kind of 3D modeling approach, which covers the whole scale of the planet-star system, has been extensively applied to a variety of hot giant exoplanets, for which the absorption in Lyα and other resonant lines were measured. Having included initially only the radiation pressure force as the ENAs generation mechanism (*Vidal-Majar et al. 2003, 2004, Lecavelier des Etangs et al. 2004, 2008, Bourrier et al. 2015*), later, these models also incorporated charge exchange process (*Holmström et al. 2008, Kislyakova et al. 2014, Bourrier et al. 2016, Lavie et al. 2017*). In particular, for the hot-neptune GJ 436b, considered in this paper, such modeling was able to reproduce not only the strongly asymmetric absorption profile and transit depth (*Ehrenreich et al 2015, Bourrier et al. 2016*), but also the observed early ingress and delayed egress of the transit light-curves (*Lavie et al. 2017*). However, despite the good agreement of numerical Monte-Carlo simulations of GJ436b with observations, the inferred best-fit parameters of PW, used as the model input, appear in obvious disagreement with the predictions of aeronomy models for this exoplanet. The chemically complex multi-species 1D model by *Loyd et al. (2017)* and 3D simulations of the escaping hydrogen-helium upper atmosphere of GJ436b by *Shaikhislamov et al. (2018b)*, both show very similar results. Specifically, the self-consistently derived in these models atmospheric mass loss rate is of about $3 \cdot 10^9$ g/s, with the outflow velocity not exceeding 10 km/s, and the maximum temperature of about 4500 K. On the other hand, the values required for the Monte-Carlo modeling in order to fit the observations, derived in *Bourrier et al. (2016)* and *Lavie et al. (2017)* are $3 \cdot 10^8$ g/s and 70 km/s, respectively. To justify such discrepancy, *Bourrier et al. (2016)* and *Lavie et al. (2017)* assumed that such a low mass loss rate is due to a very small heating efficiency by stellar XUV radiation, whereas high outflow velocity of PW is due to the acceleration by hypothetic Alfven waves.

In another recent simulation of the Lyα in-transit signatures of GJ436b with the Monte-Carlo model by *Kislyakova et al. (2019)*, somewhat different conclusions regarding the parameters of escaping PW and ENAs generation mechanism have been made. A good agreement of the modeling results with the observations both, in blue and red wings of the Lyα line, was achieved, but at the essentially smaller velocity of the escaping planetary upper atmospheric material of only ~8 km/s. Note, that such velocity value agrees well with the aeronomy simulations. Moreover, the modeling by *Kislyakova et al. (2019)* confirms the conclusion of the present paper that atomic hydrogen acceleration by the radiation pressure is insignificant for the observed in-transit Lyα signatures of GJ436b, and the key role in the production of ENAs belongs to charge-exchange.

There are several other important physical aspects that cannot be modeled with the Monte-Carlo approach and that can be adequately described only by HD/MHD models. Among those are the effects of thermal pressure and shock waves. In case of a supersonic interaction between the PW and SW, one can naturally expect the formation of shocks, or thermally compressed regions. Moreover, according to 1D, 2D, and 3D aeronomy modeling of several hot exoplanets (*Koskinen et al. 2013, Shaikhislamov et al. 2016, 2018a,b, Khodachenko et al. 2017, Loyd et al. 2017*), the PW remains collisional up to rather far distances (of several tens of $R_p$) from the planet. The exobase is located usually far beyond the Roche lobe, as well as beyond the shocked region of the interacting PW and SW. This makes the hydrodynamic approach to the modeling of plasma environments of hot exoplanets a necessary and efficient way to the understanding of physical processes there and simulation of the related transit observations.

In the present paper, a fully self-consistent 3D hydrodynamic model is applied for the first time to simulate the in-transit Lyα line absorption features of GJ436b as observed with the HST/STIS. The multi-fluid aeronomic code of the model includes the hydrogen plasma photo-chemistry, the self-consistent stellar radiation energy input to the upper atmosphere of the planet that drives its expansion, and the effects of stellar and planetary gravity, as well as the SW plasma flow. The HD/MHD modeling of hot close-orbit exoplanets has been steadily progressing nowadays from 1D to 3D codes (*Bisikalo et al. 2013, 2016, Tremblin and Chiang 2013, Trammell et al. 2014; Owen et al. 2014, Khodachenko et al. 2015, 2017, Matsakos et al. 2015, Tripathi et al. 2015, Shaikhislamov et al. 2016, 2018a,b, Erkaev et al. 2017*). At the same time, the majority of existing 3D models have not yet reached the same level of physics-and-chemistry complexity as that of the first generation of 1D aeronomy models, which would allow self-consistent simulation of the outflowing PW and its interaction with the surrounding SW. In that respect, the model, presented in this paper is the first one that includes the aeronomy part, comparable with the existing 1D simulations, while covering, at the same time, the global 3D plasma environment of the whole stellar-planetary system. It is also the only model among the existing ones, which incorporates the key processes, considered in the Monte-Carlo simulations, such as radiation pressure and charge exchange. Therefore, it can be directly compared to this kind of simulations as well. The proposed 3D hydrodynamic model not only reveals the parameters of PW, stellar XUV flux, and SW, at which good agreement of the calculated (synthetic) and experimental line absorption profiles and transit light-curves is achieved, but it also describes how different physical processes affect the observations.

Our previous paper on 3D simulation of the GJ 436b (*Shaikhislamov et al. 2018b*) was dedicated to upper atmosphere escape and did not take the stellar plasma into account. Its primary goal was to compare the results of the self-consistent 3D modeling to those of the previous 1D one and to investigate the role of helium abundance. It was found, in particular, that while the total influence of helium abundance on the mass loss is relatively small, the presence of helium affects the structuring of the nearby planetary plasmasphere. Because of its higher mass, helium component decreases the gravitational scale height of the planetary atmosphere, resulting in a sharper decrease of density with height. This shifts the $H_2$ dissociation front closer to the planet, and for GJ436b we saw two qualitatively different kinds of thermosphere – one, with an extended molecular hydrogen envelope, and another (for He/H<0.1) with a restricted $H_2$ area, closely localized around the planet. The presence of helium also affects the photo-ionization of hydrogen. While the study of helium's influence on the transit observations of GJ436b is not the major goal of the present paper, we, nevertheless, take it into account in the modeling.

It has to be noted, that the results of 3D modeling of GJ436b in *Shaikhislamov et al. (2018b)* are in a good agreement with the chemically more complex aeronomy simulation of the planet by *Loyd et al. (2017)*, which was just a 1D model, but included the oxygen and carbon components in the atmosphere. In particular, such features as temperature maximum, escaping PW velocity, $H_2$ half-dissociation height, and the mass loss rate, predicted by both models, are quantitatively similar, and the differences up to 25% can be attributed to the specifics of 3D modeling, not reproduced in 1D. The model in *Shaikhislamov et al. (2018b)* confirmed the conclusion of the previously done estimates, that the exosphere of GJ436b is indeed much extended and it obscures the most of the stellar disk, even in spite of the high impact parameter of the orbit, making the planet to transit close to the disk edge. Among the newly simulated essentially 3D features are the Coriolis twisting of the PW streams in the non-inertial planet-based reference frame and their compression towards the ecliptic plane by the stellar gravity, which in the tidally locked rotating system is not compensated by the centrifugal force in the polar directions. This effect results in the formation of a sharp plasmasphere boundaries at the distances of (10-20)·$R_p$, below and above the ecliptic plane.

The modeling results, reported in the present paper demonstrate that the huge weakly ionized plasmasphere, built around GJ436b causes the strong absorption in Lyα line, which can be as large as 70-90%. By this, the absorption at high Doppler shifted velocities is produced by the ENAs generated by charge-exchange between the planetary atoms and fast protons of the SW (in case the latter is prominently present). Therefore, the Lyα absorption is strongly asymmetric and shifted to the blue wing of the line. The interaction between the outflowing PW and SW is characterized by the presence of a clearly pronounced shocked, or compressed, region. This is the region where the ENAs are produced because the planetary atoms penetrate there through the ionopause and charge-exchange with the compressed proton fluid of the SW. The sharp early ingress starts at the position of the shock ahead of GJ436b which depends on the total pressure balance between the PW and SW. The egress, on the other hand, tends to be long-lasting in duration as a part of the escaping PW material trails behind the planet. With the dedicated model runs we determined the parameters of PW and SW, as well as the stellar XUV flux, at which good agreement of the calculated and the measured Lyα absorption profiles and the corresponding transit light-curves is achieved, and describe how different physical processes affect the observations.

It has been also found that at the Lyα radiation flux expected for GJ436, based on actual measurements and reconstruction of the line core (*Ehrenreich et al 2015*), the radiation pressure produces too small effect in the acceleration of hydrogen atoms to be seen in observations. In this respect it should be noted, that the acceleration of the escaping upper atmospheric planetary atoms by the radiation pressure has been recently reanalyzed with simple analytical estimations and hydrodynamic modeling. By this, a crucial for the acceleration process coupling between protons and hydrogen atoms, not considered in the Monte-Carlo approach, was properly taken into account (*Shaikhislamov et al. 2016, Khodachenko et al. 2017, Cherenkov et al. 2017*). In the present paper we recall the question about the role of the radiation pressure again. In Appendix section we discuss why the radiation pressure force has so little influence on the Lyα absorption features of the GJ436b.

While the modelled Lyα absorption appeared in good agreement with the measurements of major observational features, such as strong asymmetry between blue and red wings of the absorbed Lyα line, deep (>70%) in-transit depth in the high velocity blue part of the line, early ingress and extended egress parts of the transit light-curve, the exact simultaneous fitting of all observed features appeared to be difficult. The transit light-curve of GJ436b, built on the basis of all observational visits, has a complex egress part, which consists of a sharp initial drop followed by a gradual decrease, as described in Section 2. This complexity can be either a result of action of specific physical processes leading to such an ambivalent behavior of the light-curve, or appear a consequence of overlaying in one light-curve of the measurements performed at different conditions of different epochs, for which the data were obtained. The major discrepancy here is that the modeling shows the absorption maximum at 1-3 hours after the mid-transit, followed by a long egress part, which is significantly deeper than one in the observations. The egress decline can be made sharper by switching to a denser and faster SW, which blows away the trailing planetary tail. However, in this case the early ingress will be also affected, and the shallow and long part of the egress will disappear.

It is worth to note, that there are physical aspects, still not included in our model, which certainly can affect (qualitatively and quantitatively) the interaction between the PW and SW. These are first of all the magnetic fields of the star and SW on one hand, and the planetary magnetic field on the other. The presence of such large-scale (stellar and planetary) magnetic fields modifies the geometry of the regions where the ENAs are generated, thus resulting in an additional physical factor which influences the ingress and egress dynamics of the transit light-curve. Nevertheless, it should not be forgotten, that the available observational data for none of the single visits properly cover the whole transit light-curve with all its phases, i.e., the ingress, mid-transit and the extended egress parts. This fact leaves a possibility of an influence of the varying from visit to visit stellar radiation and plasma conditions on the observed behavior of the transit-light curve.

The paper is organized as follows. In Section 2, the available relevant observational data on GJ436b are briefly reviewed. Section 3 describes the model details. In Section 4, the results of simulations for the different SW conditions and XUV radiation fluxes, as well as other modeling parameters, are reported, and the role of the radiation pressure is quantitatively investigated. Section 5 presents the discussion of the obtained results and main conclusions.

## 2. Summary of the Lyα absorption measurements for GJ436b

In this Section, we briefly summarize the observational data regarding Lyα absorption during the transits of GJ436b. There are in total 8 data measurement campaigns, so-called visits, when the Lyα flux of GJ436b was recorded with *HST*, as described in (*Ehrenreich et al. 2015, Burrier et al. 2016, Lavie et al. 2017*). The visits 0, 4, 5, which cover the time interval of 26-37 hours after the mid-transit, are used to derive the out-of-transit Lyα flux of GJ436, as well as to infer its short-term and long-term variability. In-transit light-curves are constructed with data from two other sets of visits. The visits 1–3 cover about ±3.3 hours around the mid-transit and show more or less the same details of the transit light-curve. Namely, a sharp ingress, starting at about −2 hours, deep maximum of absorption, with a depth of up to 70%, continued for about 0.75 hours, and a sharp fall of at the interval of [1; 3] hours. Here and further on the phase of transit light-curve is referred as a time relative the ephemerid mid-transit time of the planet with the positive and negative values, corresponding to the phases before and after the mid-transit, respectively. The visits 6 and 7 were aimed to measure the duration of egress. They cover the interval of [4; 9] hours after the mid-transit and, as an out-of-transit reference, the interval well before the mid-transit [–8.5; –3.3]. Combined together, the visits 6 and 7 show a different transit light-curve, as compared to one revealed at the visits 1-3, which is much shallower (only 25%), but smoother and more extended in time. This difference in data can be interpreted either as an evidence of two independent transit scenarios: 1) sharp and deep; 2) long and shallow, or as a combined scenario with a sharp increase in the absorption depth during the interval [–2; 2] hours, when the planet crosses the mid-transit, followed then by a long shallow egress.

For the comparison with simulations, we use the data of all visits from *Lavie et al. (2017)*. The left panel in Figure 1 shows the transit light-curve integrated over the blue wing of Lyα line in the interval of Doppler shifted velocities [-120; -40] km/s, composed of all visits. The right panel in Figure 1 shows the absorption profiles over the whole Lyα line at the mid-transit (0 hours) and post-transit (sum of measurements at 4.2 and 5.8 hours) averaged over all visits, as well as the out-of-transit profile for the reference. The absorption depth, averaged over the blue wing interval [–120; –40] km/s of the Lyα line, reaches in some visits up to 70% of the stellar flux at the mid-transit and 32% at the post-transit. In the red wing interval [30; 110] km/s of the line it is 16% at the mid-transit and 22% at the post-transit, respectively. Thus, the Lyα absorption of GJ436b at the mid-transit is very asymmetric, being much stronger in the blue wing of the line, while in the post-transit it is more symmetric and reaches moderate values in both, the blue and red wing intervals. These features of the Lyα absorption are related with the dynamical conditions realized in the interacting PW and SW around GJ434b. The detailed modeling of this interaction and the related distribution and motion of the absorbing hydrogen sheds light on the key physical processes in the system, transit scenarios, and dominating absorption mechanisms.

Further on, for the purpose of comparison of the modeled and observed transit light-curves and the Lyα absorption profiles we use the same observational data as those displayed in Figure 1, if not specified otherwise.

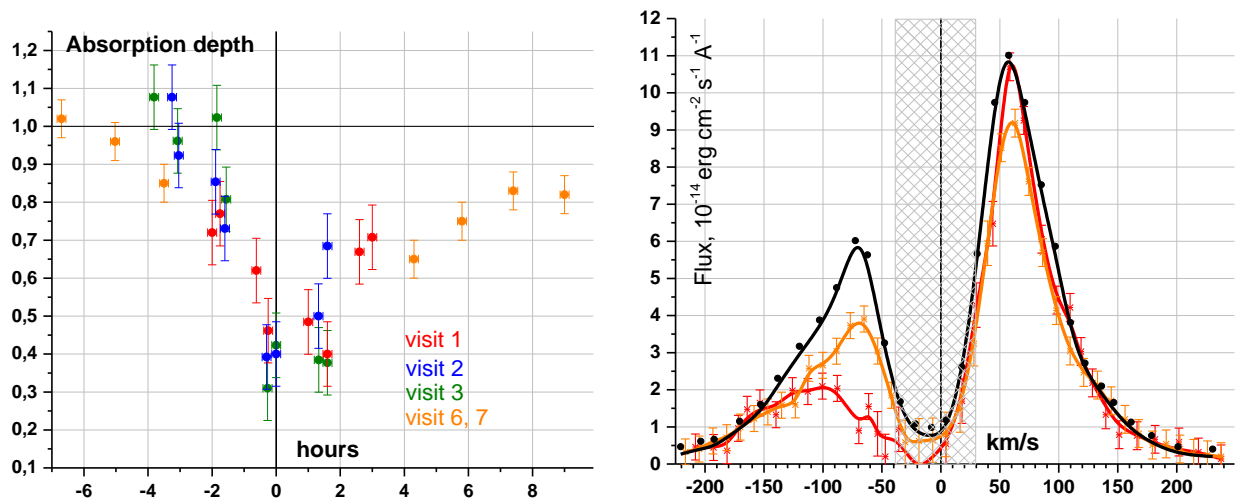

**Figure 1.** *Left panel:* in-transit Lyα light-curve of GJ436b integrated over the blue wing [-120; -40] km/s of the line. *Right panel:* Spectral profiles of the Lyα line averaged over all visits for the out-of-transit (black), mid-transit (red), and post-transit (orange). Data are reproduced from *Lavie et al. (2017),* where further details can be found. The checked region of [-40; 30] km/s indicates the interval of geocoronal contamination of measurements.

## 3. The Model

The used here 3D multi-fluid model further extends the 3D model of GJ436b, presented in *Shaikhislamov et al. (2018b)*. The code solves numerically the hydrodynamic equations of continuity, momentum, and energy for all species (marked with index *j*) of the simulated multi-component flow, written in the following form *(Shaikhislamov et al. 2016)*:

$$\frac{\partial}{\partial t} n_j + \nabla (\mathbf{V}_j n_j) = N_{XUV,j} + N_{exh,j} \tag{3.1}$$

$$m_j \frac{\partial}{\partial t} \mathbf{V}_j + m (\mathbf{V}_j \nabla) \mathbf{V}_j = -\frac{1}{n_j} \nabla n_j k T_j - \frac{z_j}{n_e} \nabla n_e k T_e - m_j \nabla U - 2 m_j \mathbf{V}_j \times \mathbf{\Omega} - m_j \sum_i C_{ji}^v (\mathbf{V}_j - \mathbf{V}_i) \tag{3.2}$$

$$\frac{\partial}{\partial t} T_j + (\mathbf{V}_j \nabla) T_j + (\gamma - 1) T_j \nabla \mathbf{V}_j = W_{XUV} - \sum_i C_{ji}^T (T_j - T_i) - W_{cool} \tag{3.3}$$

Among the considered fluid species are the hydrogen atoms (H), ions ($H^+$), and electrons of the planetary and stellar origin. The atomic hydrogen component is not originally present in the SW. It is produced by charge exchange between the SW protons and planetary atomic hydrogen. Account of molecular hydrogen ($H_2$) and the corresponding molecular ions $H_2^+$, $H_3^+$, like e.g. in *Shaikhislamov et al. (2018a,b)* and *Khodachenko et al. (2017)*, enables more accurate treatment of the inner regions of the planetary thermosphere. The model admits also the inclusion of the minor species, which are described as separate fluids by the corresponding momentum and the ionization-recombination equations. In the present modeling of GJ436b, besides of the dominating hydrogen, we consider also helium (He) component, taken with an abundance He/H=0.1 (if not specified otherwise).

The basic physical parameters of the model and simulation results are scaled in units of the characteristic values of the problem: temperature, in $T_0 = 10^4$ K, speed, in ion thermal speed, $V_0 = V_{Ti} = 9.07$ km s$^{-1}$, distance, in radius $R_p=0.25 \cdot 10^{10}$ cm of GJ436b, and time, in $\tau = R_p/V_0 \sim 0.75$ hour.

The main processes, responsible for the transformation between neutral and ionized particles are photo-ionization, electron impact ionization and dielectronic recombination. These are included in the term $N_{XUV,j}$ in the continuity equation (3.1) and are applied for all species. Photo-ionization also results in a strong heating of the planetary atmospheric material by the produced photo-electrons. The corresponding term $W_{XUV,j}$ in the energy equation (3.3) (addressed in *Trammell et al. 2011, Shaikhislamov et al. 2014, Khodachenko et al. 2015*) is derived by integration of the stellar XUV spectrum. For the M-dwarf GJ436 we use the XUV spectrum compiled by the MUSCLES survey *(France et al. 2016)* which is based on actually measured fluxes in VUV and X-ray bands. The corresponding integrated XUV flux (λ<91.2 nm) is about $F_{XUV}$=0.86 erg s$^{-1}$ cm$^{-2}$ at 1 a.u. We also applied the regression model by *Arkhypov et al. (2018)* based on the statistical analysis of available stellar parameters (rotation periods and effective temperatures) and observed activity rates, which predicts for GJ436 the flux $F_{XUV} \sim 2$ erg cm$^{-2}$ s$^{-1}$ at 1 a.u. Note, that the same values of 1–2 erg cm$^{-2}$ s$^{-1}$ at 1 a.u. are estimated also in *Bourrier et al. (2016)* and *Lavie et al. (2017)*. The XUV photons ionize hydrogen atoms, as well as $H_2$ and He components according to the wavelength dependent cross-sections. The model assumes that the energy, released in the form of photo-electrons is rapidly and equally re-distributed between all locally present particles with an efficiency of 50%. This is a commonly used assumption, which we adopted on the basis of qualitative analysis *(Shaikhislamov et al. 2014)*. The attenuation of the XUV flux inside the planetary atmosphere is calculated for each spectral bin according to the wavelength dependent cross-sections. The heating term includes also energy losses due to excitation and ionization of hydrogen atoms, and in a simplified form it can be written as follows:

$$W_{XUV} = \frac{1}{N_{tot}} (\gamma - 1) \cdot n_a \left[ \langle (\hbar \nu - E_{ion}) \sigma_{XUV} F_{XUV} \rangle - n_e \upsilon_{Te} (E_{21} \sigma_{12} + E_{ion} \sigma_{ion}) \right] \quad (3.4)$$

Of certain importance are also the energy losses due to infrared radiation of $H_3^+$ molecule with a rate $C_{H3+}$ calculated by *Miller at al. (2013)*, and due to dissociation of $H_2$ molecule, $W_{cool} = (\gamma - 1) \cdot (n_{H3+} \cdot C_{H3+} + n_{H2} \cdot E_{diss} \cdot C_{diss})/N_{tot}$. Here, $N_{tot}$ is the total number density of all particles, including electrons.

Another kind of exchange between the considered particle populations are the resonant charge-exchange collisions. Indeed, charge-exchange has a typical cross-section of $\sigma_{exc} = 6 \cdot 10^{-15} \text{cm}^2$ at low energies, which is an order of magnitude larger than the elastic collision cross-section. Charge-exchange process is included in the term $N_{exh,j}$. Also, when the planetary atoms and protons slip relative each other, because they have different thermal pressure profiles, and the protons additionally feel the electron pressure while the atoms do not, the charge-exchange between them leads to their velocity and temperature interchange. We describe this process with a collision rate coefficient $C_{ji}^\upsilon$, where the upper index indicates the physical quantity being interchanged. For example, in the momentum equation for planetary protons it reads as $C_{H^+H}^\upsilon = n_H^{pw} \sigma_{exh} \upsilon$, where the relative velocity $\upsilon \approx \sqrt{V_{Ti}^2 + V_{Tj}^2 + (V_j - V_i)^2}$ depends in general on thermal speeds and relative velocities of the interacting fluids (in the considered example – of protons and hydrogen atoms). Besides the charge-exchange, ordinary elastic and Coulomb collisions are included in $C_{ji}^\upsilon$ and $C_{ji}^T$ terms.

For the typical parameters of the planetary plasmasphere, Coulomb collisions effectively couple the ionized species between each other and with electrons. For example, at $T<10^4$ K and $n_{H+} > 10^6 \text{cm}^{-3}$ the collisional equalization time for temperature and momentum $\left( C_{H+,H+}^\upsilon \right)^{-1} = \tau_{Coul} \approx 2.4 \cdot T^{1.5}(K)/n_{H+}$ is less than 1–10 s for protons (*Braginskii, 1965*). This is several orders of magnitude less than the typical gasdynamic time scale of the problem $\tau = R_p/V_0 \sim 3 \cdot 10^3$ s. Therefore, there is no need to calculate the dynamics of every charged component of the considered medium, and we assume in the present simulations all of them (i.e., $H^+, H_2^+, H_3^+$, $He^+$) to have the same temperature and velocity. On the other hand, the temperature and velocity of each neutral component (i.e., H, $H_2$, He) are calculated individually.

The model equations are solved in a non-inertial spherical frame of reference fixed at the planet center, which orbits together with the planet and rotates with the same rate Ω so, that one of the axes is continuously pointed to the star. The Z-axis is directed perpendicular to the ecliptic plane. This is a so-called tidally locked frame of reference. Though, in general case the planet itself can rotate around its axis with an own (different) rate, in this paper, GJ436b is taken to be also tidally locked to the star. In this tidally locked frame of reference we properly account the non-inertial terms, i.e., the generalized gravity potential and Coriolis force. The particular expressions for these terms are well known, and they were given, e.g., in *Khodachenko et al. (2015)* and Shaikhislamov et al. (2016). The fluid velocity at the planetary surface is taken to be zero, whereas at the stellar surface, the normal component of the plasma flow velocity is defined by the polytropic solution of the SW model, described below. The tangential component of SW velocity is connected with the stellar surface rotation in the chosen non-inertial frame of reference. It is worth to note, that those velocities at the stellar surface are in fact very small, as compared to the characteristic velocity scale $V_0$ of the problem. To keep the number of points operated in the numerical code, tractable for processing, the radial mesh is highly non-uniform, with the grid step increasing exponentially from the planet surface. This allows resolving of the highly stratified upper atmosphere of the planet, where the required grid step is as small as $\Delta r = R_p/400$. For the analysis of the simulation results, a related Cartesian coordinate system with the X-axis, pointing from the planet to the star, is also used further. Further details on the numerical implementation of the code are given in Appendix.

For the self-consistent calculations on the scale of the whole star-planet system, we incorporate in the same code the SW plasma dynamics as well. Besides the upper planetary atmosphere, the stellar corona appears as another boundary of the simulation domain, at which the corresponding coronal values of the plasma parameters are fixed. Outside this boundary, taken for simplicity at the stellar surface with the radius $R_{star}$, the SW, i.e. the proton fluid, is calculated by the same code as one used for the protons of PW. The simulation of SW is a complex problem, which has been extensively (in itself extensively studied) tackled in the last decades (see e.g., *Usmanov et al. 2011* and references therein). Its detailed treatment, especially the processes of the SW heating and acceleration up to super-sonic velocities (still not yet fully understood), appears beyond the scope of the present paper. A common approach of the majority of the global astrophysical codes, adopted for the simulation of exoplanetary environments, is to model the SW with a simplified approach, using a polytropic or even an isothermal specific heat ratio $1 \leq \gamma_p \leq 1.2$ (e.g., in *Matsakos at al 2015, Bisikalo et al. 2013, Christie et al. 2016, Daley-Yates & Stevens 2018*). Any polytropic index, lower than the adiabatic one $\gamma_a = 5/3$, introduces in the energy equation an effective source term which replaces the variety of complex SW energizing processes. However, simulation with a non-adiabatic index makes unrealistic the treatment of shocks, developed in the region of the colliding SW and PW flows. To avoid of this difficulty, we introduce in energy equation an empirical heating source to drive the SW. The corresponding source term is found from a semi-analytical solution of 1D polytropic Parker-like model, which yields for a given stellar gravity, the base temperature $T_{cor}$, and the asymptotic supersonic velocity $V_{sw,\infty}$, a unique value of $\gamma_p$ and the radial profiles of all SW parameters (see, e.g., in *Keppens & Goedbloed 1999*). To fix the SW plasma density at the stellar surface boundary, we use an integral characteristic parameter – stellar mass loss rate $M'_{sw}$. The heating term, therefore, is found as:

$$W_{sw}(R) = (\gamma_a - \gamma_p) \cdot T_p(R) \cdot divV_p(R) \qquad (3.5)$$

Here, $T_p$ and $V_p$ are obtained from the polytropic solution, and the distance R is measured from the center of star. It was specially checked in the simulations that without PW, the solution obtained for SW is in a good agreement with the analytical model. Note, that the equation (3.5) specifies a spatially distributed, continuous and stationary in time heat source, which is defined by the specified above model parameters only. To avoid of unphysical artefacts, the heating source is switched off in the regions where a significant amount of particles of the planetary origin penetrate into the SW. The modeling approach, described above, allows a physically self-consistent global simulation of the generation of super-sonic PW and SW, as well as their interaction.

For the sake of generality and for comparison with other models, we also include in the code the radiation pressure force, acting on the hydrogen atoms due to the Lyα flux. The reconstructed Lyα line profile, based on the actual measurements for GJ436 can be found in *Bourrier et al.*

*(2015)*. The total flux of $\approx 1$ erg cm$^2$s$^{-1}$ at the reference distance of 1 a.u. generates at the center of the emission line a force, equal to $\approx 0.7$ of the stellar gravity pull. The radial dependence of the Lyα flux and the particle instant velocity relative the line center are properly taken into account during the calculation of the acceleration of the hydrogen atoms. An important factor is the Lyα flux self-shielding. This we compute by Voigt convolution of the Lorentz line shape with the natural width and the Maxwellian distribution of atoms with the temperature and density calculated by the hydrodynamic model. The convolution integral is expressed via the empirical formula from *Tasitsiomi (2006)*, as described in *Shaikhislamov et al. (2016)* and *Khodachenko et al. (2017)*. The self-shielding is calculated in the code in 30 km/s bins in the range of $\pm 120$ km/s.

The synthetic absorption in the Lyα line by the transiting GJ436b is calculated with a dedicated data processing program, on the basis of the numerical simulation results, as follows. We use the available observational data on the spectrally resolved out-of-transit and in-transit fluxes, $F_{out}(\lambda)$ and $F_{in}(\lambda)$, respectively. By this, the integrated absorption in a particular part $[\lambda_1, \lambda_2]$ of the line profile is calculated as

$$\bar{A}_{abs}(\lambda_1, \lambda_2) = 1 - \int_{\lambda_1}^{\lambda_2} F_{in} d\lambda \Big/ \int_{\lambda_1}^{\lambda_2} F_{out} d\lambda. \tag{3.6}$$

The physical parameters of the dynamic planetary environment, obtained in course of the simulations are used to calculate the absorption $A_{abs}(\lambda)$ of a given atomic line (in the present study – of the Lyα line) in dependence on λ, while the emission line profile it-self is not specified. Since, by definition $F_{in}(\lambda) = [1 - A_{abs}(\lambda)] \cdot F_{out}(\lambda)$, the integrated synthetic absorption is obtained, using equation (3.6), by averaging of $A_{abs}(\lambda)$ over the line emission profile $F_{out}(\lambda)$, measured in observations:

$$\bar{A}_{abs,sim}(\lambda_1, \lambda_2) = 1 - \int_{\lambda_1}^{\lambda_2} [1 - A_{abs}(\lambda)] \cdot F_{out} d\lambda \Big/ \int_{\lambda_1}^{\lambda_2} F_{out} d\lambda \tag{3.7}$$

If not specified otherwise, we compare the simulated synthetic transit light-curves $\bar{A}_{abs,sim}(\lambda_1, \lambda_2, t)$ with those derived from the observations e.g., in *Ehrenreich et al. (2015), Bourrier et al. (2016), Lavie et al. (2017)* in the interval $[\lambda_1, \lambda_2]$, corresponding to the Doppler blue-shifted [-120; -40] and red-shifted [30; 110] km/s velocities. The time *t* is counted relative to the optical mid-transit of the planet.

**Table 1.** Parameters of GJ436 system used in the simulations.

| Planet: | GJ 436b |
|---|---|
| **Planetary mass:** | $0.07 M_J$ |
| **Planetary radius:** | $0.35 R_J$ |
| **Orbital distance:** | 0.029 a.u. |
| **Period:** | 2.64 days |
| **Inclination:** | 86.4 degrees |
| **Stellar mass:** | $0.45 M_{Sun}$ |
| **Stellar radius:** | $0.46 R_{Sun}$ |

A summary of the parameters of the star-planet system, used in the modeling is given in Table 1. Further on, the following parameters, mostly based on the analogy with the Sun and general physical reasoning, will be assumed as the so-called *"standard"* ones: upper atmospheric temperature and pressure of GJ436b, i.e. at the inner boundary of the simulation domain (at the conventional planet surface) $T_{base}$=750 K and $P_{base}$=0.05 bar, respectively; helium abundance, He/H=0.1; ionizing radiation of GJ436 at 1 a.u., $F_{XUV}$=0.86 erg cm$^{-2}$ s$^{-1}$; stellar coronal temperature $T_{cor}$=2·10$^6$ K, the terminal SW velocity $V_{sw,\infty}$=400 km/s, and the stellar mass loss rate $M'_{sw}$=2.5·10$^{11}$ g/s. Note, that the assumed stellar radiation and mass loss rate are about 5-10 times less than those for the Sun, which is mostly due to the smaller size of GJ436.

## 4. Simulation results

About 100 simulations were performed with the varying of most important parameters of the system in relatively broad ranges, specifically, with: the stellar XUV flux scaled to 1 a.u., $F_{xuv}$ =(0.4–5.0) erg cm$^{-2}$ s$^{-1}$; the terminal SW velocity, $V_{sw,\infty}$=(150–2000) km/s; and the stellar mass loss rate, $M'_{sw}$=(10$^{10}$–10$^{13}$) g/s. The total mass loss of the star is an unconstrained parameter, so it was varied from the small values, at which SW does not affect the planetary atmospheric material outflow, up to extremely large figures, exceeding those of the quiet Sun more than an order of magnitude.

Besides these main parameters of the system, there are several others, that to a lesser extent also affect the absorption in Lyα. Among those is the stellar coronal temperature, $T_{cor}$, which may influence the formation of the shock, and which was varied for different modeling runs in the range from 1.5 to 4.5 MK. Also, the changes of cooling rate due to $H_3^+$ molecule were investigated, which being varied by an order of magnitude, affects the PW intensity up to ~50%. Note, that this cooling rate so far was not directly measured. It is based on a complicated theoretical calculations and estimates (*Miller et al. 2013*). With a series of dedicated simulation runs, it has been also found, that variation by a factor of two of such parameters of the upper planetary atmosphere (i.e., of the inner boundary conditions) as $T_{base}$, $P_{base}$, and helium abundance, does not produce any significant difference in the modeling results.

### 4.1 Weak stellar wind and the role of radiation pressure force

First, we consider the so-called '*captured by the star*' regime of the PW and SW interaction (*Shaikhislamov et al. 2016*). It corresponds to the case of a sufficiently weak SW, when the escaping PW material is captured by the stellar gravity and accumulates around the star. For the *standard* parameters of the system specified above, the *captured by the star* regime takes place for a small stellar mass loss rate $M'_{sw}$~5·10$^9$ g/s, which is ~500 times less than that of the present Sun. Note, that at four times higher $M'_{sw}$=2·10$^{10}$ g/s, the SW is already strong enough (i.e. dense and fast) to blow away the escaping PW material, which corresponds to the '*blown by the wind*' regime of the PW and SW interaction according the terminology, used in *Shaikhislamov et al.( 2016)*. Figure 2 shows the color plots of H and H+ density distributions and the typical streamlines, achieved after about 25 orbital revolutions of the planet after the start of the simulation run.

One can see that the PW stream spirals and quickly (in several tens revolutions) falls on the star. The calculation of long term accretion and(or) accumulation of the material around the star requires the inclusion of more sophisticated physics than that in the present version of the code. The performed simulations show that the *captured by the star* regime of the PW and SW interaction is unlikely for GJ436b, because under the typical standard parameters of the system it takes place only at extremely low stellar mass loss rates. Therefore, in further modeling and to study of the Lyα absorption features of GJ436b only the *blown by the wind* regime is considered. At the same time, the *captured by the star* regime is the most suitable one for the demonstration of the role of radiation pressure.

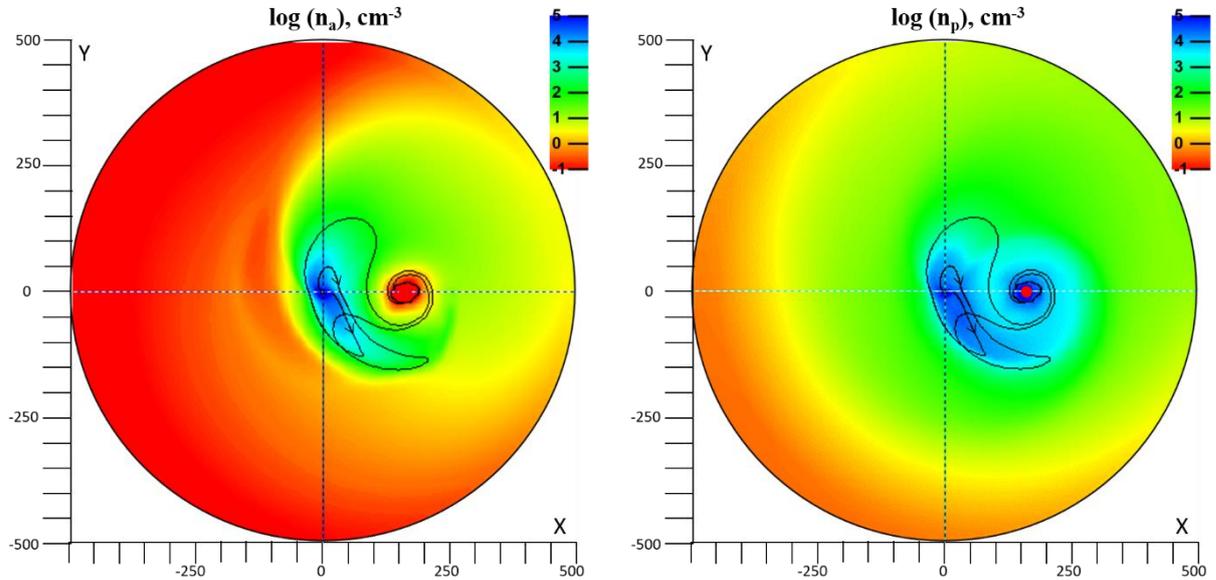

**Figure 2.** Distribution of atomic hydrogen (*left panel*) and proton (*right panel*) densities in the equatorial plane (X-Y), calculated with the *standard* parameters of the model: $F_{XUV}$=0.85 erg s$^{-1}$ cm$^{-2}$ at 1 a.u., $T_{base}$=750 K, $P_{base}$=0.05 bar, He/H=0.1, $T_{cor}$=2·10$^6$ K, $V_{sw,\infty}$=4·10$^7$ cm/s, $\dot{M}_{sw}$=5·10$^9$ g/s. The distance is scaled in units of GJ436b radius, $R_p$. The planet is located at the center of the coordinate system. The star is located to the right from the planet at X=158$R_p$ and is marked in the right panel with the scaled red circle. Black lines in both panels show the streamlines of the material flow.

Left panel in Figure 3 shows the same plots as those in Figure 2 (left panel), but obtained with the inclusion of the radiation pressure force, produced by the stellar Lyα flux of $F_{Ly\alpha}$=2 erg cm$^{-2}$ s$^{-1}$ (scaled to 1 a.u.). At this value the acceleration of hydrogen atoms due to the radiation pressure force is about the same as that, caused by the stellar gravity. Although, the PW material also accumulates around the star as in previous case, one can clearly see that the hydrogen atoms are pushed a bit further away from the star and the spiraling of the PW stream lasts significantly longer than in the simulations without radiation pressure. The extended helmet-shape structure, seen in the meridional cut across the ecliptic plane in the right panel in Figure 3 is produced mainly due to the radiation pressure force, and it is practically absent without it. The latter is clearly indicated also by the difference between green halos ($n_a$ density level ~10$^2$ cm$^{-3}$) in the ecliptic plane in the left panels of Figures 2 and 3.

For the quantitative characterization of the role of the radiation pressure force in terms of its measurable manifestation in the Lyα absorption profiles, these, as well as the corresponding transit light-curves, were calculated for both cases (i.e. with and without inclusion of the radiation pressure). The results are shown in Figure 4. In the case of weak SW and without the radiation pressure force, the absorption in the high velocity blue wing of the Lyα line is just a few percent (as expected). Note, that besides of the spike at the mid-transit, the absorption in this case takes place over an extended time interval before and after the transit. This happens due to formation of an extended belt of atomic hydrogen around the star supplied mainly by the escaping PW material. The inclusion of radiation pressure force does not make significant difference, although additional acceleration of atoms results in a pushing of the accreting PW stream away from the star. The consequences of this effect can be seen in the left panel of Figure 4 before the transit (i.e., ahead of the planet). The Lyα absorption profiles provided in the right panel of Figure 4 show that the velocities of hydrogen atoms are shifted, due to the action of the radiation pressure force, on about –20 km/s.

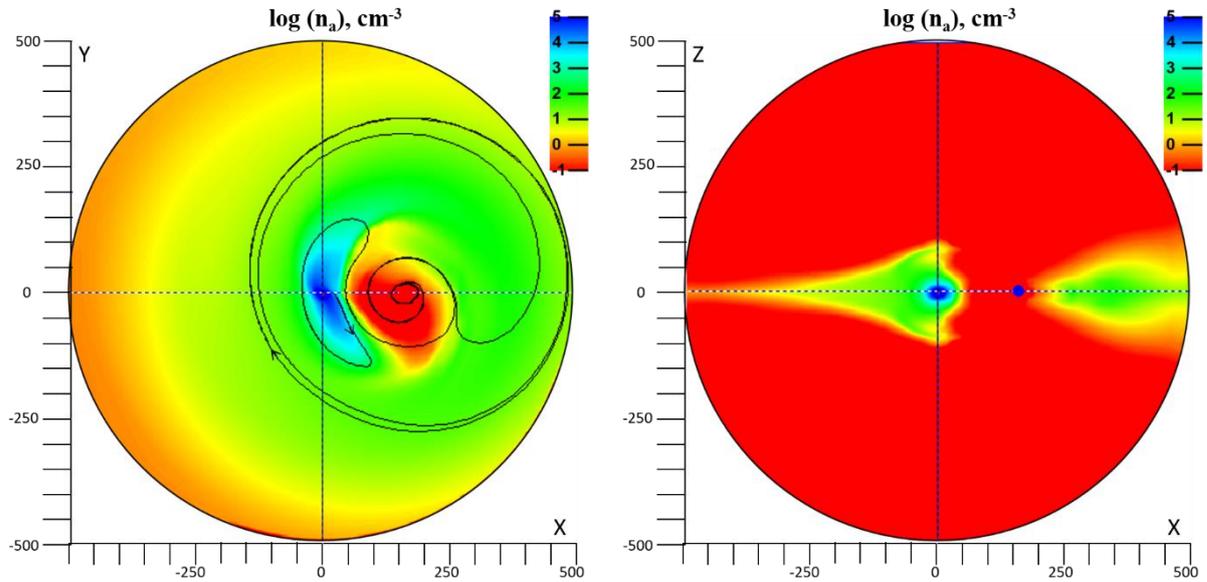

**Figure 3.** Distribution of atomic hydrogen density in the equatorial (*X-Y*) (*left panel*) and meridional (*X-Z*) (*right panel*) planes, calculated with the account of the radiation pressure force due to the Lyα flux of 2 erg s$^{-1}$ cm$^{-2}$ at 1 a.u., while keeping other parameters the same as those in Figure 2. Black lines show the streamlines of the neutral material flow.

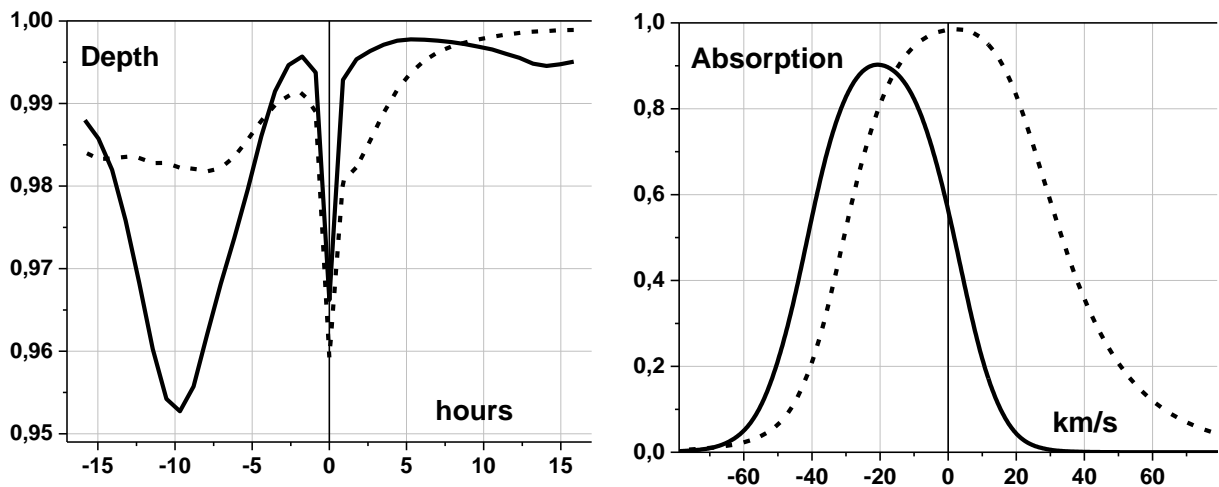

**Figure 4.** *Left panel:* The simulated transit light-curves in the blue wing of the Lyα line averaged over the Doppler shifted velocities [−120; −40] km/s. *Right panel:* absorption profile around the Lyα line center at the transit phase t= −10 hours. Dashed and solid lines correspond to the cases without (also in Figure 2) and with (also in Figure 3) the inclusion of the radiation pressure force, respectively.

The effect of atomic hydrogen acceleration by the radiation pressure force is further demonstrated in Figure 5 in the velocity (radial component) profiles along the planet-star line. As it can be seen, if the radiation pressure is not included, the hydrogen atoms move towards the star, while being accelerated by the stellar gravity up to velocities of ~100 km/s. The radiation pressure force, as expected from the estimations, nearly balances the stellar gravity, so that the hydrogen atoms do not fall to the star, except of the small region of R<50 where atoms are coupled to the protons, and a complex mixing of the PW and SW material takes place. Note also that far from the planet, where the material is rarified and hydrogen atoms become uncoupled from the protons, the streamlines of hydrogen atoms (see in Figure 3, left panel) are almost circular, as being additionally supported by the radiation pressure force, atoms do not fall towards the star.

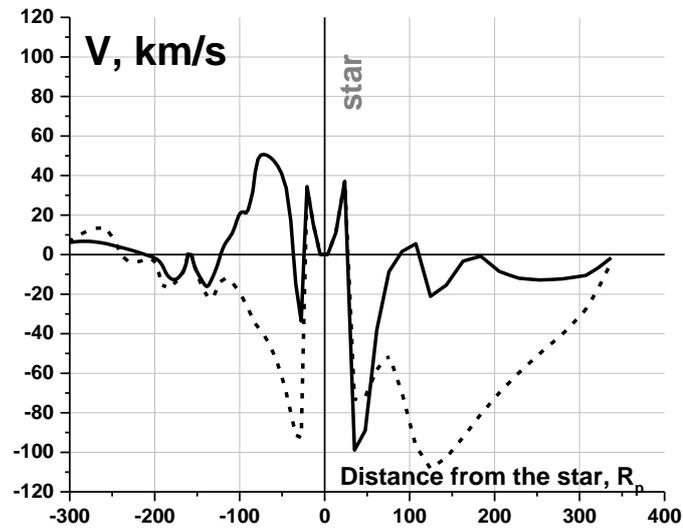

**Figure 5.** Velocity (radial component) of hydrogen atoms along the planet-star line (Z=0, Y=0) in the star-centered coordinate system. The planet is located at X= −158. Positive/negative values of the velocity mean the motion away/towards the star. Dashed and solid lines correspond to the cases without (also in Figure 2) and with (also in Figure 3) the inclusion of the radiation pressure force, respectively.

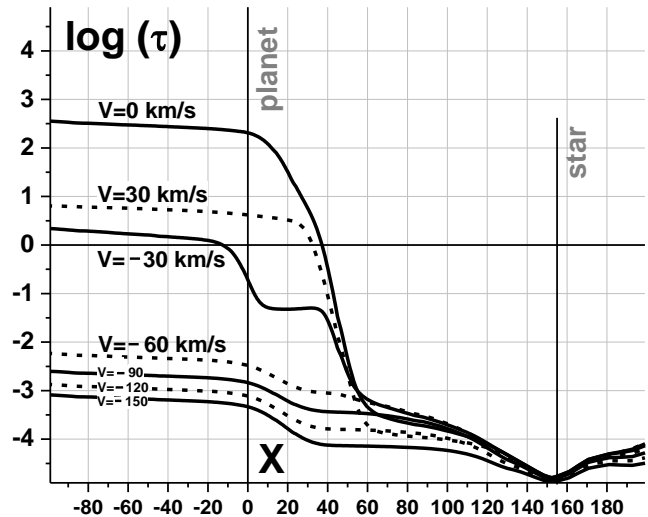

**Figure 6.** The logarithm of the integral optical depth for the Lyα photons versus distance from the planet along the line-of-sight parallel to X-axis, for a set of different Doppler shifted velocities. Negative velocity values correspond to the motion away from the star. The integration goes along the X axis, at fixed Y=−20 and Z=0, starting from the star position at X=158.

Finally, one of the reasons why the effect of the radiation pressure force is not prominent in the case of GJ436b under study is the self-shielding. Figure 6 shows the optical depth for the Lyα photons $\tau(V) = \int_{X_{star}}^{X} dx \cdot n_H \cdot \sigma_{abs}(V, V_x, T)$ as a function of distance along the line-of-sight, passing parallel to X-axis and sufficiently close (Y=−20, Z=0) to the planet, for a set of different Doppler shifted velocities $V = c \cdot (1 - \lambda/\lambda_0)$. As it can be concluded from the plots in Figure 6, the central part of the Lyα line in the range [–30; 30] km/s is strongly absorbed in the close vicinity of the planet, e.g., for |Y|<20. This means that close to the planet the acceleration of slow atoms (<30km/s) cannot take place because of the Lyα shielding.

Altogether, the numerical simulations and tests, performed with the inclusion of all the key physics, demonstrate that the radiation pressure force produces a relatively small effect on the Lyα absorption for GJ436b. Therefore, for simplicity's sake, it will not be included in further modeling calculations. Nevertheless, to stay on the safe side with this kind of simplification, in several cases considered below, we continued the comparison of the simulation results with and without the radiation pressure force, finding always a very little difference between them.

The question on the role of the radiation pressure force in the Lyα absorption by GJ 436b is further discussed in Appendix 1, where the results of numerical simulations are supported and justified with a theoretical analysis and analytic estimates.

## 4.2 Material escape on GJ436b; interaction with SW – modeling under different conditions

The parameters of the discussed further simulation runs are listed in Table 2. For the referencing purpose, a certain number (1..14) is assigned to each set of parameters. Figures 7 and 8 show the color plots of atomic hydrogen and proton density distributions, obtained in the modeling run with the parameter set No.1. They reveal that the SW flow sweeps away the escaping PW material and redirects it to the trailing tail. A compressed layer is formed in front of the planet, well seen in the in the ENAs and temperature distributions (Figures 7 and 8, right panels) as well as in the temperature plot along the planet-star line in Figure 9.

| No. | XUV | $M'_{pw}$ | $V_{sw,\infty}$ km/s | $T_{cor}$ MK | $M'_{sw}$ | $V_{sw,pl}$ km/s | $T_{swp}$ MK | $n_{swp}$ cm$^{-3}$ | $D_{blue\,Max}$ | $D_{blue}$ t=0 | $D_{blue}$ t=5 h | $D_{red}$ t=0 | $D_{red}$ t=5 h | other |
|---|---|---|---|---|---|---|---|---|---|---|---|---|---|---|
| 1 | 0.86 | 0.2 | 400 | 2 | 25 | 170 | 0.6 | 4000 | 68 | 58% | 43% | 14% | 0% | |
| 2 | 0.43 | 0.1 | 400 | 2 | 25 | 170 | 0.6 | 4000 | 60 | 45 | 30 | 8 | 0 | |
| 3 | 1.7 | 0.4 | 400 | 2 | 25 | 170 | 0.6 | 4000 | 62 | 49 | 47 | 18 | 5 | |
| 4 | 0.86 | 0.2 | 400 | 2 | 12.5 | 170 | 0.6 | 2000 | 57 | 43 | 48 | 15 | 7.5 | |
| 5 | 0.86 | 0.2 | 400 | 2 | 50 | 170 | 0.6 | 8000 | 69 | 45 | 31 | 9 | 0 | |
| 6 | 0.86 | 0.2 | 400 | 2 | 7.5 | 170 | 0.6 | 1200 | 44 | 25 | 40 | 15 | 2 | |
| 7 | 0.86 | 0.21 | 400 | 200 | 25 | 170 | 0.6 | 4000 | 34 | 24 | 14 | 5.4 | 0 | He/H=1 |
| 8 | 0.86 | 0.21 | 400 | 200 | 25 | 170 | 0.6 | 4000 | 57 | 45 | 29 | 13 | 0 | He/H=0.2 |
| 9 | 0.86 | 0.29 | 400 | 200 | 25 | 170 | 0.6 | 4000 | 78 | 65 | 57 | 21.5 | 1 | He/H=0.05 |
| 10 | 0.86 | 0.29 | 400 | 200 | 25 | 170 | 0.6 | 4000 | 84 | 74 | 68 | 21.5 | 1 | He/H=10$^{-3}$ |
| 11 | 3.0 | 0.76 | 1100 | 4 | 14 | 600 | 2.2 | 450 | 73 | 50 | 5 | 17 | 0 | |
| 12 | 3.0 | 0.89 | 720 | 3.5 | 34 | 450 | 1.3 | 1500 | 82 | 80 | 6 | 19 | 0 | *$^1$ |
| 13 | 1.3 | 0.36 | 220 | 2.8 | 12 | 100 | 0.35 | 2700 | 29 | 25 | 27 | 23 | 11 | |
| 14 | | | | | | | | | 66 | 66 | 32 | 16 | 22 | *$^2$ |

**Table 2.** The parameter sets for the discussed simulation. XUV column contains the stellar flux in the range of wavelengths 1<λ<91.2 nm, expressed in erg cm$^{-2}$ s$^{-1}$ and scaled to 1 a.u. The mass loss rates of the planet $M'_{pw}$ and the star $M'_{sw}$ are expressed in 10$^{10}$ g/s. If not specified otherwise, other parameters of the simulations are: $P_{base}$=0.05 bar, $T_{base}$=750 K, He/H=0.1. Index "swp" denotes SW parameters at the orbit of the planet. The $D_{blue\,Max}$ column contains the maximum absorption depth values in the blue wing interval [−120; −40]. The $D_{blue\,t=0}$, $D_{red\,t=0}$, $D_{blue\,t=5h}$, and $D_{red\,t=5h}$ columns contain the absorption depth values at the mid-transit (t=0 h) and post-transit (t=5 h) in the blue wing [−120; −40] and red wing [30; 110] km/s intervals, respectively. The last (right) column contains the remarks on other modeling parameters varied.
*$^1$: 5 times decreased cooling rate by H3$^+$ molecule.
*$^2$: Absorption values obtained in observations.

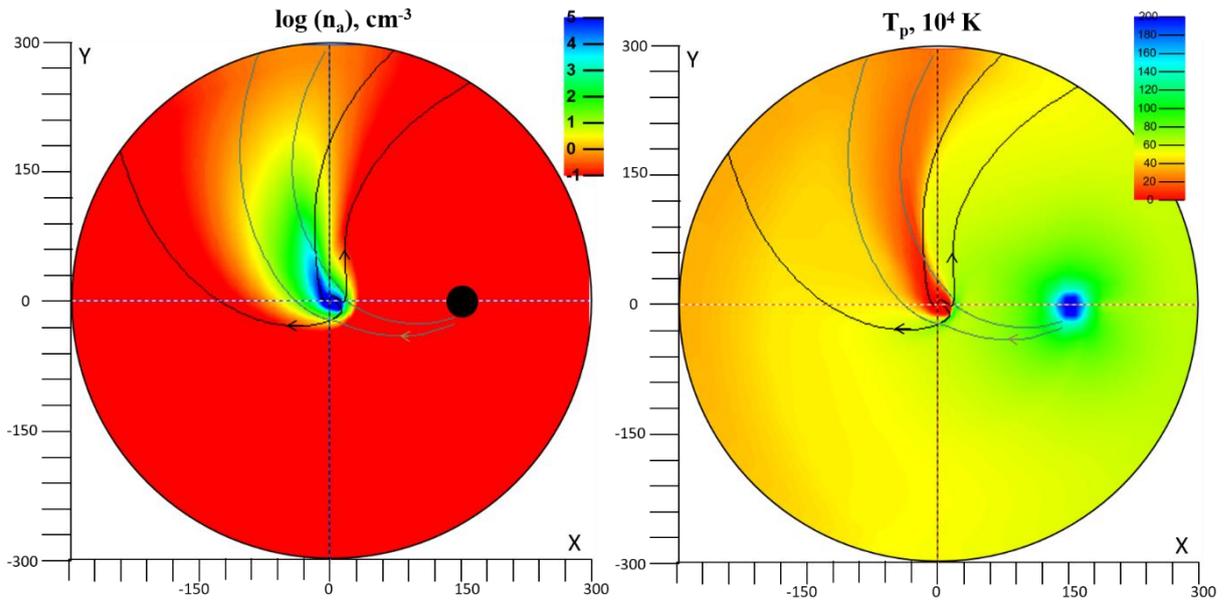

**Figure 7.** Distribution of atomic hydrogen density (*left panel*) and proton temperature (*right panel*) in the equatorial plane (X-Y), calculated with the parameter set No.1. Black and khaki streamlines show the flow of hydrogen atoms and protons, respectively.

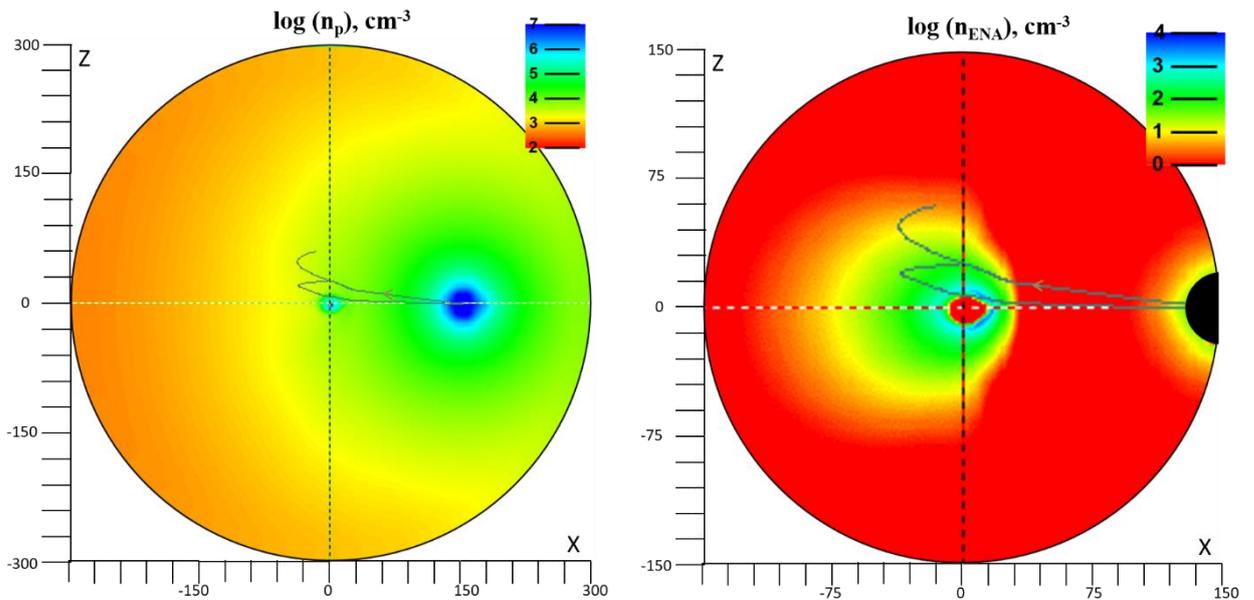

**Figure 8.** Distribution of proton (*left panel*) and ENA (*right panel*, twice zoomed) densities in the meridional (X-Z) plane, calculated with the parameter set No.1. Khaki lines in both panels show the streamlines of the proton flow.

However, due to the high temperature of SW plasma, the relative SW and PW velocity is not enough to generate a strong shock wave (the thermal speed in the SW plasma at the orbit of planet is ~130 km/s). The presence of a weak shock is seen in Figures 7 and 8 as a slight discontinuity of the proton streamlines at the bowshock. The ENAs are mainly generated in the region between the ionopause (~$20R_p$) and the bowshock (~$40R_p$), both clearly seen in Figure 8 (right panel), as sharp boundaries in the ENAs density color plot. The neutral atoms of planetary origin, which approach closely to the ionopause, penetrate into the shocked region where they become uncoupled from the protons. The structure of different regions along the planet-star line (i.e. at Z=0, Y=0), from the planetary upper atmosphere up to the star, is shown in more details in Figure 9. All these features have been already addressed and discussed with respect to our previous 2D simulations (*Shaikhislamov et al. 2016, Khodachenko et al. 2017*). They are qualitatively similar to those, revealed in 2D case during the "captured by the star" regime of SW and PW interaction, though quantitatively there are differences, related with the specifics of 3D, which cannot be reproduced in 2D modeling.

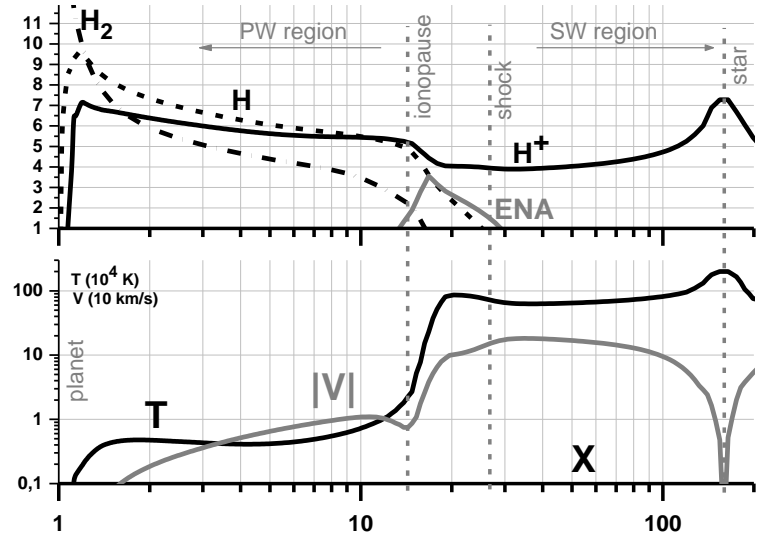

**Figure 9.** Distribution of the key physical parameters of the GJ436b environment along X-axis, at Z=0, Y=0 (i.e., the planet-star), simulated with the parameter set No.1. *Upper panel*: density (in $cm^{-3}$; log scale) of protons (H+, solid black line), molecular hydrogen ($H_2$, dotted line), atomic hydrogen (H, dashed line) and ENAs (solid gray line); *Lower panel*: proton temperature (black line) and velocity (gray line).

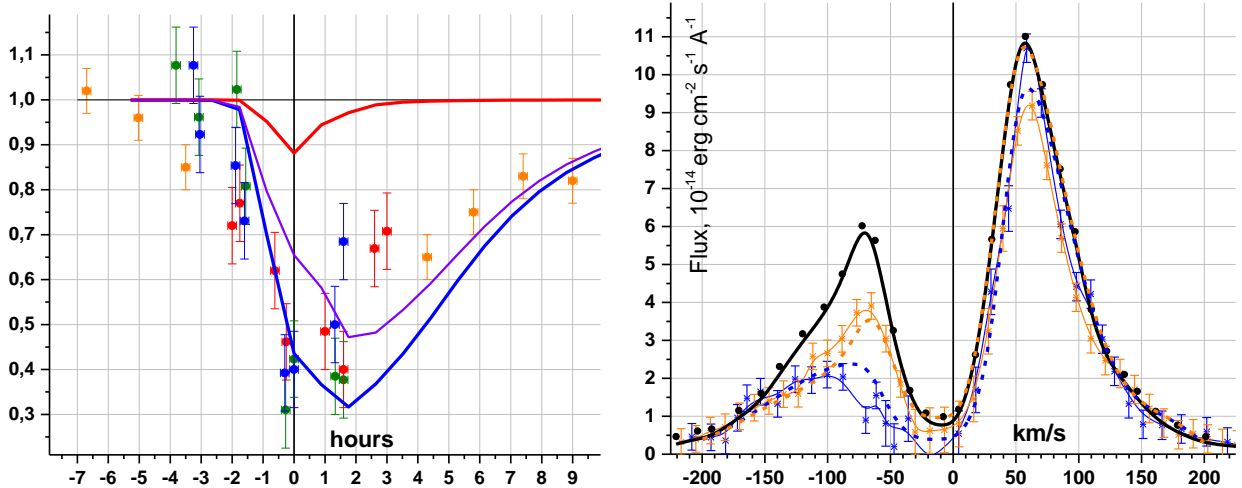

**Figure 10.** *Left panel*: The transit light-curves of GJ436b in the blue ([-120; -40] km/s; blue line) and red ([30; 110] km/s; red line) wings of the Lyα line, simulated with the parameter set No.1. Thin violet line shows the part of the Lyα absorption in the blue wing caused by the ENAs. Observed data are shown by squares with error bars. *Right panel*: the corresponding modeled Lyα line profiles at the mid-transit (t=0 h; blue dots) and post-transit (t=5 h; orange dots) phases. The measured line profiles are shown with blue and orange stars with error bars and with black circles, for the same mid-transit and post-transit phases, respectively.

Next, we describe the absorption in Lyα derived in the simulation run with the parameter set No.1. Figure 10 shows the transit light-curve in the blue wing of the Lyα line and the modeled full line profiles at the mid-transit (t=0 h) and post-transit (t=5 h), respectively. For comparison, the measured data from Figure 1 (according *Lavie et al. 2017*) are reproduced as well.

As it can be seen in Figure 10, our self-consistent 3D modeling of the dynamical environment of GJ436b reproduces the main observational feature, namely, the strong absorption in the blue wing of the Lyα line. Under the conditions of the modeling parameter set No.1, it reaches almost 70%. We note that these parameters have not been specifically fitted, but rather assumed as the most relevant and typical ones. The simulated Lyα absorption is strongly asymmetric, and in the red wing of the line it does not exceed 15%. This is because the absorption at velocities >30 km/s is to a significant extend due to the ENAs, created by charge-exchange between the planetary atoms and SW protons. This fact is illustrated by the relative closeness of the blue and violet lines in the left panel of Figure 10. Note, that the modeled transit light-curve possesses quite an extended egress part that gradually decreases for as long as 10 hours after the mid-transit. This duration corresponds to about 1/6 of the whole orbit of GJ436b. Such effect is caused by the long trailing planetary tail. The line profiles in the right panel of Figure 10 show that the absorption takes place mostly in the blue wing and extends up to −150 km/s, which is in good agreement with the observations.

A view of the geometry of the Lyα absorbing region is presented in Figure 11, where the color map of the line-of-sight absorption, expressed via the optical depth as $1-\exp(-\tau)$ and averaged over the blue wing of the Lyα line, is shown. Integration of this distribution over the whole stellar disk gives the absorption ≈60%, which corresponds the mid-transit point (t=0 h) on the transit light-curve in Figure 10 (left panel).

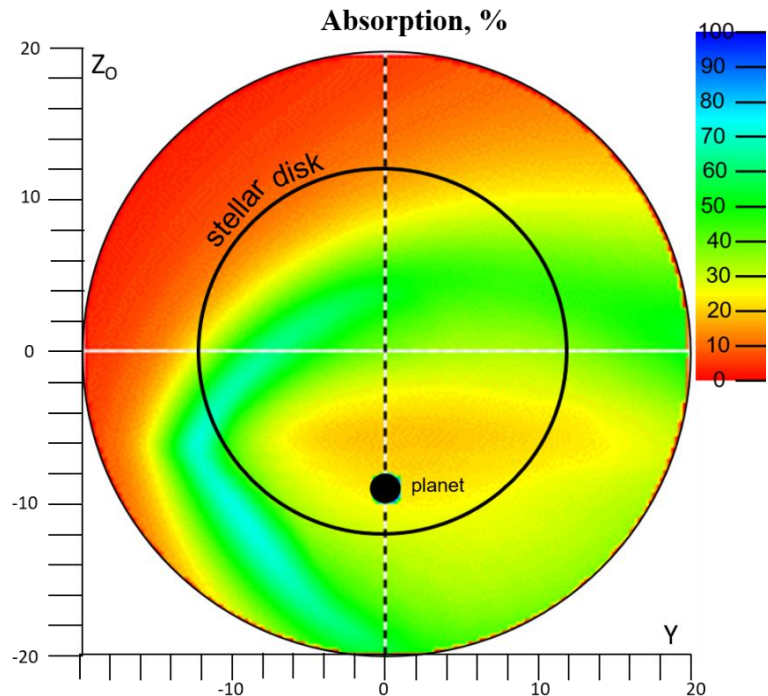

**Figure 11.** Distribution of the line-of-sight absorption, averaged over the blue wing of the Lyα line, as seen by remote observer at the mid-transit. Stellar disk is shown with a black-line circumference, and GJ436b, with a black circle. Note, that this picture uses the observer reference frame, in which the planet is shifted to $Z_O$=-9 relative the observer-star line ($Z_O$=0, Y=0).

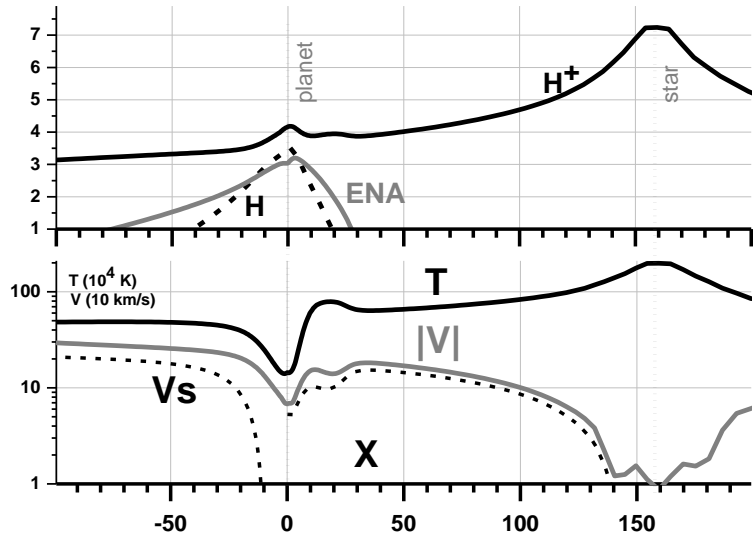

**Figure 12.** Distribution profiles of the key physical parameters of the environment of GJ436b along X-axis, at Z=11, Y=0, simulated with the parameter set No.1. *Upper panel*: density (in cm$^{-3}$; log scale) of protons (H+, solid black line), atomic hydrogen (H, dashed line), and ENAs (solid gray line); *Lower panel*: proton temperature (black line), absolute velocity (gray line), and the component of velocity along the line-of-sight (dotted line).

One can clearly see in Figure 11 that the strongest absorption comes from the regions associated with the shock, located relatively far from the planet (up to ~11R$_p$). The compressed layer between the planetary ionopause and bowshock, where the ENAs are generated, covers the most of central part of the stellar disk. Because of that, the grazing trajectory of the planet, as seen by an observer (i.e., from the Earth), is actually favorable for the absorption. If the planet would cross the stellar disk close to the star's center, the maximum absorption would be only 43%, and the egress part of the transit light-curve, twice shorter in duration. The details of the plasma parameters' distribution are shown in Figure 12, where the profiles of proton, atomic hydrogen, and ENA densities along the line-of-sight, parallel to X-axis are presented for the fixed Z=11, and Y=0 (the planet-centered coordinate system is used here). One can see that along this particular line-of-sight, which in the observer reference frame in Figure 11 has the coordinates $Z_O=2$, $Y_O=0$, the major part of absorption should happen in the interval of X= [−30; 20], mostly populated by the atomic hydrogen fraction. Since the considered line-of-sight crosses the shocked region, the corresponding temperature and velocity there are relatively high. Figure 12 also shows the component of velocity, Vs, along the line-of-sight, which is positive everywhere, except a small region near the planet. It confirms that the neutrals move away from the star with the speeds, corresponding to the Doppler shifted velocities of the blue wing of the Lyα line, from −10 km/s to −200 km/s.

Next series of simulation runs with the parameter sets No.1-6 demonstrates how the modeling results are changed due to a factor of two variation of the major two parameters of the problem, specifically, the stellar XUV flux $F_{XUV}$ and the mass loss rate of the star $M'_{sw}$. Note, that the intensity of SW (i.e. the mass flux) in all these cases is varied by changing of the SW density, while keeping its velocity and temperature the same. The corresponding transit light-curves and absorption profiles are shown in Figure 13. Since the maxima of the simulated transit light-curves are shifted relative the ephemerid center of transit, the absorption profiles in the right panel of Figure 13 are shown for the time t=1 h, i.e., slightly after the mid-transit. One can see that the decrease/increase of the XUV flux and SW intensity directly influences the starting time of the early ingress and the duration of egress. The maximum depth depends mostly on the density of SW, which influences the efficiency of ENAs generation. In general, the variation of

$F_{XUV}$ and $M'_{sw}$ by a factor of two results in variation of the absorption in the blue wing of the Lyα line by a factor of 1.5, while the red wing remains unaffected.

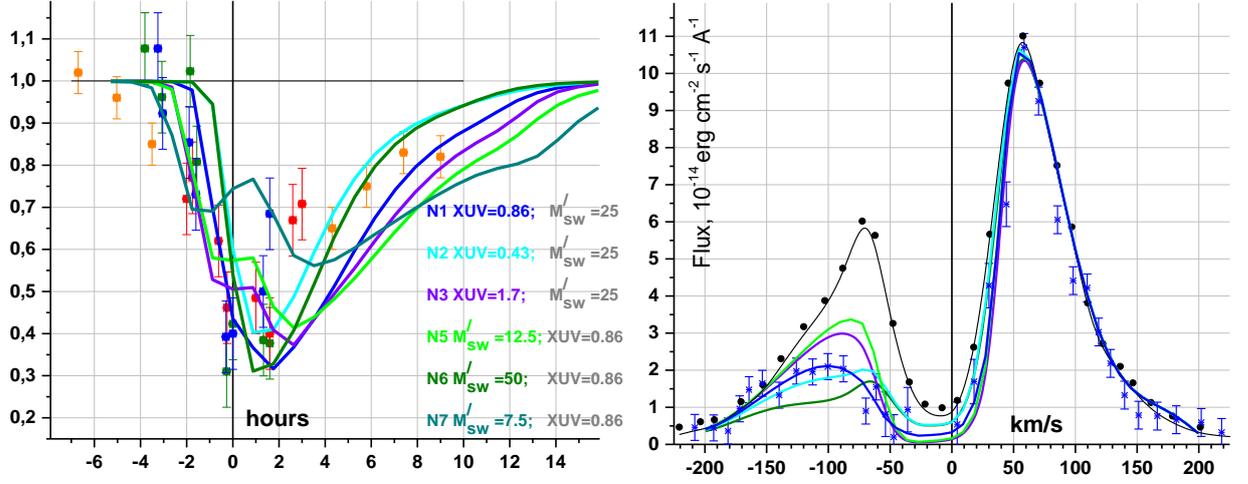

**Figure 13.** *Left panel*: the transit light-curves of GJ 436b in the blue ([-120; -40] km/s) wing of Lyα line, simulated with the parameter sets No.1-6, i.e. for different XUV fluxes (0.43; 0.86; 1.7 erg cm$^{-2}$ s$^{-1}$ at 1 a.u.) and SW intensities defined via the integral mass loss rate (7.5; 12.5; 25; 50 in units of $10^{10}$ g/s). *Right panel*: the corresponding modeled Lyα line profiles at t=1 h (i.e., near the mid-transit) the for parameter sets No.1-5, coded with the same colors as in the left panel. The measured line profiles out of transit and at the mid-transit (t=0 h) are shown with black circles and blue stars, respectively.

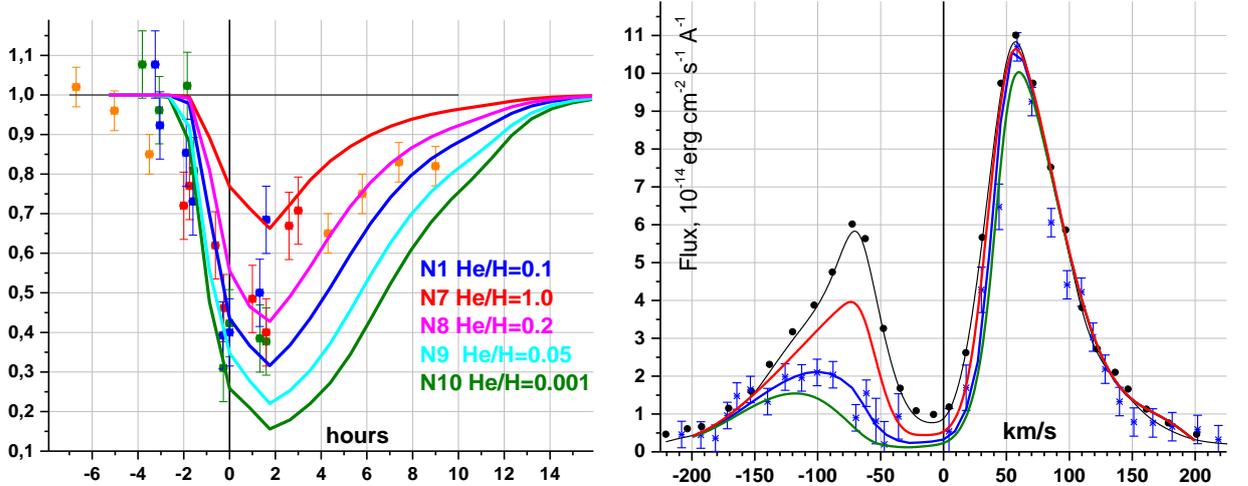

**Figure 14.** *Left panel*: the transit light-curves of GJ436b in the blue ([-120; -40] km/s) wing of Lyα line, simulated with the parameter sets No.1 and No.7-10, i.e., with different helium abundances (He/H=0.001; 0.05; 0.1; 0.2; 1). *Right panel*: the corresponding modeled Lyα line profiles at t=1 h (i.e., near the mid-transit) for the parameter sets No.1,7,10, coded with the same colors as in the left panel. The measured line profiles out of transit and at the mid-transit (t=0 h) are shown with black circles and blue stars, respectively.

Another parameter which affects the absorption is the helium abundance He/H. When it is very small, the PW outflow velocity is higher, resulting in a denser hydrogen cloud around the planet (*Shaikhislamov et al. 2018b*) and an extremely deep and broad transit light-curve. Figure 14 shows the simulated transit light-curves of GJ436b in the blue wing of the Lyα line and the corresponding modeled Lyα line profiles at t=1 h (i.e., close to the mid-transit), obtained for different helium abundances with the model parameter sets No.1, and No.7-10. When He/H increases to about unity, the PW velocity significantly decreases because of the mass loading by heavier helium atoms, resulting in less hydrogen outflow and, correspondingly, less absorption in Lyα.

Despite the general good agreement between the modeling results and observations, some of the features, obtained in the simulations and shown above, do not fit the observations in all details. In particular, the ingress starts ~1 h later; the maximum of absorption is also shifted by ~1 h; the post-transit phase of the light-curve looks like a simple decline, instead of the two-stage decreasing with a sharp fall, followed by a slow decay phase. Also, the depth of egress in general is much higher. While in the simulations with the parameter set No.1, the absorption profile at the mid-transit in the blue wing of the Lyα line is in a good general agreement with the observations i.e., quantitative differences are not crucial (see in Figure 10), in the red wing there is a noticeable qualitative difference. The modeled absorption in the Doppler shift velocities interval [30; 110] km/s at the mid-transit is larger, than at the post-transit, while in the observations the situation is opposite.

The specially performed modeling study reveals that it is difficult to fit all the features of the observed Lyα absorption just by varying of the available modeling parameters. The start-time of the ingress is easily shifted by a factor of two variation of the SW pressure (e.g., by manipulation with the SW density, and to certain extend, with the velocity). It can be also adjusted by modifying of the planetary mass loss rate via the variation of the stellar XUV flux (see e.g. in Figure 13). However, fitting of the early ingress observations leads to an extended egress, resulting in an even stronger disagreement with the observations in the post-transit part of the light-curve. On the other hand, the egress fall time can be adjusted by increasing of the SW velocity and pressure. In this case, the ingress starts too late, and the absorption in the red wing of the Lyα line becomes too small. The absorption in the red wing, in its turn, is sensitive to the heating of planetary atoms due to the interaction with SW and it can be increased by decreasing of the SW velocity together with increasing of its temperature. The post-transit absorption in the red wing of the Lyα line at the level of about 20% is achieved in the simulations under the conditions of a rather small SW velocity, $V_{swp}$<100 km/s, at the planet orbit.

In order to fit the transit light-curves and the absorption profiles more precisely, checking of other physical parameters and possible processes, was done. Specifically, under consideration appeared 1) the heating efficiency by the XUV radiation, which influences the intensity of the PW outflow, while keeping the photo-ionization the same; 2) the cooling rate due to H3+, which affects the thermosphere structure; 3) the base temperature and density, which in a small degree control the mass loss rate; and 4) the planet rotation rate around its axis, which to a certain degree affects the zonal flows in the thermosphere. However, the resulting manifestations of the considered parameter variations were found to be either small, or not suitable for the intended fitting.

As it was already pointed out in Section 2, the available observations can be interpreted as two different scenarios, realized under different conditions. Below we consider them both in more details. The *first scenario* with a sharp ingress, starting at –2 h, and deep mid-transit ~70% phase, lasting for about 2 hours, is followed by a sharp egress. In course of the performed simulation runs, it has been found that such kind of the transit light-curve can be realized under the conditions of a very fast SW with an asymptotic terminal velocity reaching the values of $V_{sw,\infty}$ = 700…2000 km/s. At such velocities, the spiraling of material flow, due to Coriolis force, in the planet-fixed frame of reference is less significant, resulting in a more pronounced sweeping of the PW tail. This can be seen in Figure 15, obtained with the modeling parameter set No.12. In this case, the inclination of the planetary tail to the orbit trajectory is about 32$^o$, whereas under the conditions of parameter set No.1 (for comparison), it is it is only 16$^o$

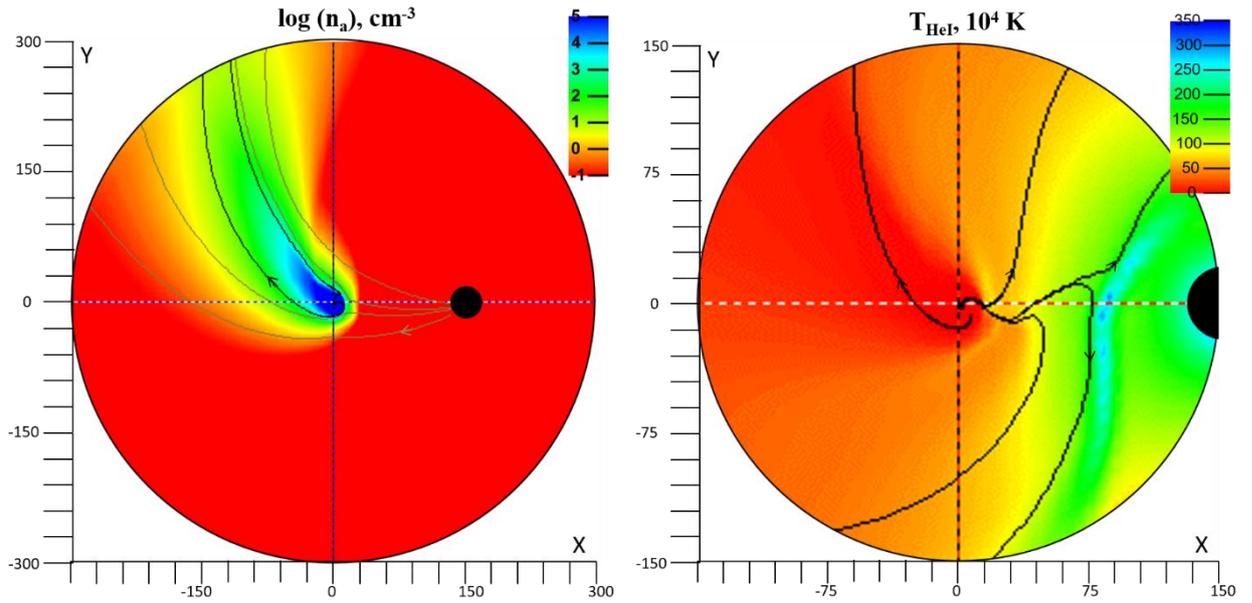

**Figure 15.** Distribution of density of hydrogen atoms (*left panel*) and temperature of helium atoms *(right panel)* in the equatorial plane (X-Y), calculated with the parameter set No.12. Black and khaki streamlines show the flow of neutrals (HI – in the left panel, and HeI – in the right panel) and protons, respectively.

As demonstrated in Figure 16, obtained in the simulations with the modeling parameter sets No.11 and 12 (i.e., fast SW and high $F_{XUV}$), a lesser inclination of the planetary tail relative the line-of-sight results in a significantly sharper post-transit decline of the light-curve. To compensate the higher pressure of the fast SW (proportional to the SW temperature and squared velocity) and to reproduce the effect of early ingress, one has to increase the pressure of PW as well. This is why the higher values of XUV flux are taken in the applied simulation parameter sets No.11 and No.12. Note, that the Ly$\alpha$ line profile simulated with the parameter set No.11 at the mid-transit nicely fits the observations in the blue wing and remains also in good agreement with the integral depth in the red wing.

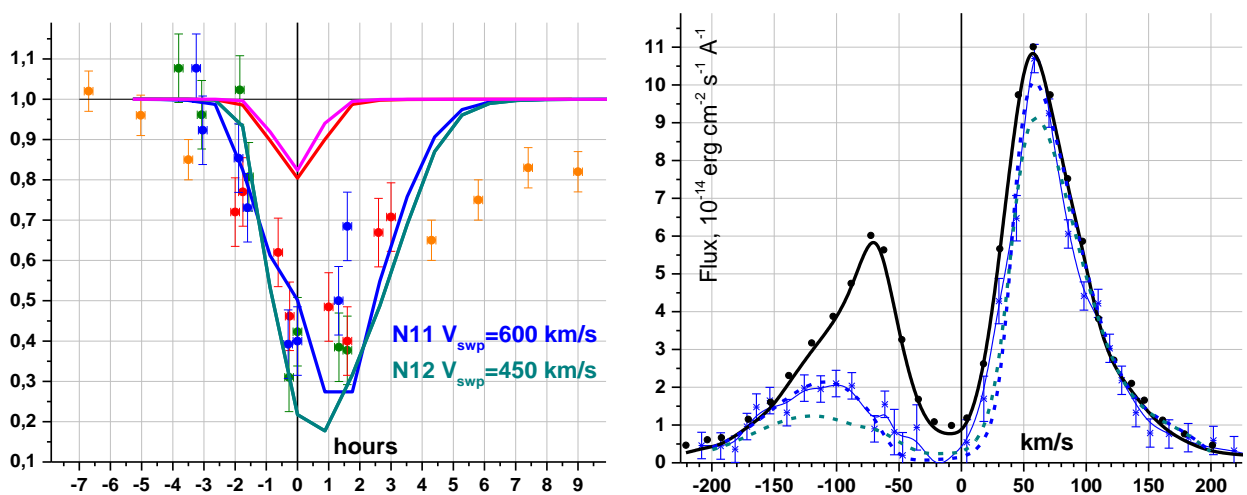

**Figure 16.** *Left panel*: the transit light-curves of GJ436b in the blue ([-120; -40] km/s) and red ([30; 110] km/s) wings of the Ly$\alpha$ line, simulated with the parameter sets No.11 (blue and red lines) and No.12 (cyan and magenta lines). *Right panel*: the corresponding modeled Ly$\alpha$ line profiles at the mid-transit (t=0 h) for the simulations with the parameter sets No.11 (blue dots) and No.12 (cyan dots), respectively. The measured line profiles out of transit and at the mid-transit (t=0 h) are shown with black circles and blue stars, respectively.

The right panel in Figure 15 shows the features which do not affect the observations, but demonstrate the capacity of the multi-fluid model to reproduce the fine details of the complex PW and SW interaction. Specifically, the neutral particles, such as helium and hydrogen atoms, are coupled to ions in the PW flow, but decouple beyond the ionopause, in the shocked region, where the density of the stellar protons is not high enough for the collisional mixing. As demonstrated in the right panel of Figure 15 for helium, as example, in these regions the temperature of planetary neutrals is distinctly different from the temperature of SW. Closer to the star, the proton density becomes again sufficiently high for elastic collisions to stop helium atoms falling on the star. For a supersonic helium flow this results in the generation of a shock, positioned at the distance of ~60$R_p$ from the star. At this distance, the density of atomic helium is orders of magnitude lower than that of the SW protons and the shock does not influence the parameters of SW. Correspondingly, the shock generated in the atomic hydrogen fluid, is positioned farther from the star, because hydrogen atoms and protons intermix each other via charge-exchange, which has larger cross-section than the elastic collisions.

The *second scenario*, with an extended and shallow ingress and egress phases and a relatively small maximal depth of ~25%, is suggested for the transits, revealed in the visits 6 and 7 (*Lavie at al. 2017*). In our simulations such scenario was realized under the conditions of slow SW with a low pressure. The slow SW velocity is needed to increase the absorption in the red wing of the Lyα line, while the low pressure enables expanding of the PW stream ahead the planet, resulting in an earlier ingress. To compensate the overall decrease of the absorption due to the low density of SW and the related less amount of the generated ENAs, a slightly increased XUV flux is necessary. The transit light-curves and Lyα line absorption profile at the post-transit, obtained with the modeling parameter set No.13, are shown in Figure 17. One can see that the transit starts sufficiently early, which is in agreement with the data from visits 6 and 7, and lasts for a long time of ~20 hours at approximately the same level of ~25%. The simulated Lyα absorption profile in the red wing agrees with the measured profile in the range [40; 65] km/s, though it differs at larger velocities.

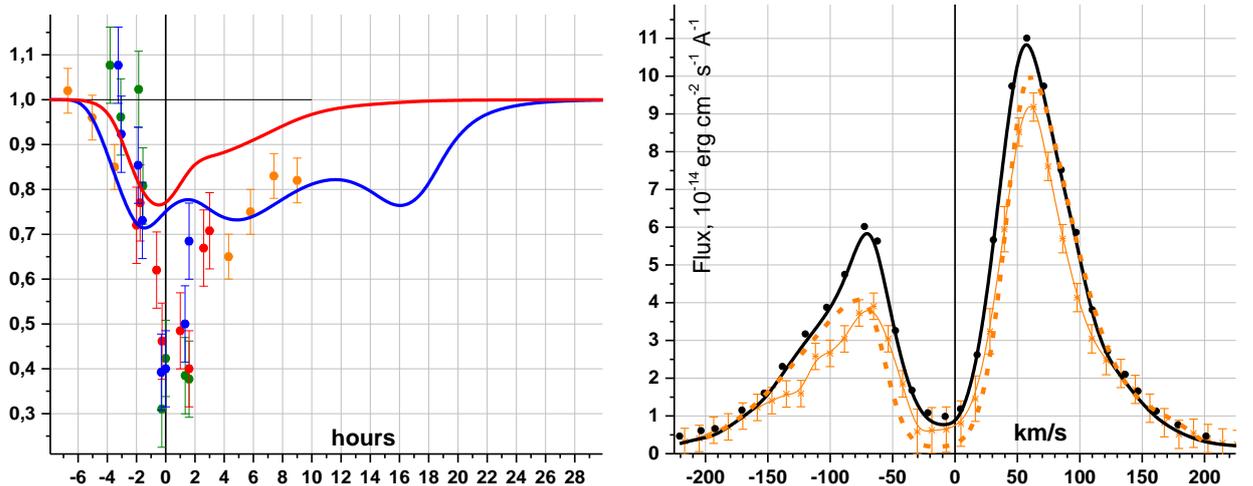

**Figure 17.** *Left panel*: the transit light-curves of GJ436b in the blue ([-120; -40] km/s; blue line) and red ([30; 110] km/s; red line) wings of the Lyα line, simulated with the parameter set No.13. *Right panel*: the corresponding modeled Lyα line profiles at the post-transit phase (t=5 h, orange dots). The measured line profiles out-of-transit and at the post-transit are shown with black circles and orange stars, respectively.

## 5. Discussion and Conclusions

Let's estimate how high is the absorption by the expanded upper atmosphere of GJ436b inside the Roche lobe, due to the natural line broadening mechanism. At the velocity of 60 km/s, the corresponding cross-section of the Lyα line is about $4\cdot 10^{-19}$ cm$^2$. Taking the typical density $n_H$ ~$10^7$ cm$^{-3}$ (see e.g., in Figure 9), the optical depth can be estimated as follows: $\tau \sim n_{HI}\sigma_{nat}R_{Lag1} \sim 0.06$, where $R_{Lag1}=5.8R_p$ is the distance to the 1$^{st}$ Lagrange point. The integral absorption should be further reduced by a factor of $(R_{Lag1}/R_{star})^2$. Thus, as expected, the natural line broadening is too small for GJ436b to play any significant role. The thermal broadening for the resonant atoms has a much larger cross-section, $6\cdot 10^{-14}\cdot\sqrt{10^4 K/T}$ cm$^2$. As it follows from the simulations in Figure 12, the bulk velocity $V_{bulk}$ in the shocked region varies in the range [−10; −150] km/s, whereas the thermal velocity of hot hydrogen atoms with the Maxwellian distribution at the temperature T~$4\cdot 10^5$ K has a comparable value $V_{thermal}\sim V_{bulk}$. The integrated column density of ENAs and planetary atoms is about NL~$6\cdot 10^2\cdot 50\cdot R_p \sim 10^{14}$ cm$^{-2}$. Thus, the optical depth for the thermal line broadening absorption is about unity, $NL\sigma_{res}$ ~1. Since the distributions of physical parameters, shown in Figure 12 for the particularly chosen line-of-sight, almost do not change over the significant part of the integration region, as it can be seen in Figure 11, the absorption due to the thermal broadening mechanism should be significant in the blue wing of the Lyα line, in agreement with the observations.

A typical product of the PW and SW interaction is the formation of a bowshock, or a thermally compressed region ahead of the planet, whereas the escaping PW material trails the planet as a tail, due to momentum conservation. The pressure exerted by the SW pushes the tail away from the star. The propagation of PW material above and below the ecliptic plane is restricted by the stellar gravity. One of the distinct features of the observed transit light-curves is an early ingress, which starts at about t= −2 hours. With respect to the planet, in view of its orbital motion, this corresponds to the position of about $30\cdot R_p$ ahead. The supersonic interaction of PW with SW naturally creates such a sharp boundary in the form of ionopause and compression layer/bowshock. The position (so-called stand-off distance) of the ionopause is determined by the pressure balance condition between the PW and SW. To estimate it, one can assume that the largest part of the SW total pressure is contributed by the dynamic pressure component related with the orbital motion of GJ436b, because the background SW at the orbital distance of ~0.028 a.u. is still less than the Keplerian speed and is approximately perpendicular to the direction of the orbital motion. Therefore, expressing the total pressure of PW via the planetary mass loss rate as follows:

$$p_{sw} \approx m_p n_{sw} V_{orb}^2 \sim p_{pw} \approx m_p n_{pw}(V_{pw}^2 + kT_{pw}/m_p) \approx M'_{pw}(V_{pw} + kT_{pw}/m_p V_{pw})/(4\pi r^2),$$

one can obtain the ionopause stand-off distance:

$$r_{ionopause}/R_p \sim \sqrt{\frac{M'_{pw}(V_{pw}+kT_{pw}/m_p V_{pw})}{4\pi R_p^2 m_p V_{orb}^2 n_{sw}}} \quad . \tag{4.1}$$

In particular, for $M'_{pw}=3\cdot 10^9$ g/s, $V_{pw}=5$ km/s, $T_{pw}=10^4$ K, and $n_{sw}=4\cdot 10^3$ cm$^{-3}$, the ionopause stand-off distance (i.e. the position of pressure balance point) is $r_{ionopause}/R_p$ ~10. The bowshock should be located 1.5–2 times farther, i.e., at $r_{shock}/R_p$ ~20. These estimates approximately agree with the observed timing of the early ingress.

While the early ingress in the Lyα absorption can be explained by the bowshock formation, the sharp decrease of the transit light-curve which takes place in 1.5 hours after the mid-transit, cannot. Due to the conservation of the orbital momentum and the low speed of the escaping PW, as compared to the orbital velocity, the tail-ward flow of the PW should approximately follow the planetary orbit, producing long egress in the absorption. The sharp drop in the absorption also cannot be explained by the photo-ionization of atoms, as the corresponding ionization time exceeds 10 hours. One of the factors, that might help cutting the egress, is the dynamic pressure of SW. If the velocity of SW exceeds significantly the orbital speed, then the tail of the escaping PW would be swept by SW from the planet, extending more in the direction of the star-planet line, rather than being trailed along the orbital trajectory. In this case, the duration of the absorption at the egress phase would be determined by the width of the PW stream. However, the performed simulations show that a very high SW velocity ~1000 km/s is needed for such a scenario to take place, which in case of the Sun is achieved only in the coronal mass ejections (CMEs). It has to be noted that reducing of the egress duration by means of increasing of the SW density, instead of velocity, results in a simultaneous cut of the early ingress part, which would contradict the observations. Another observational feature, detected in the visits 1 and 6, which is not reproduced in simulations, is the absorption in the red wing of the Lyα line at relatively high velocities [60; 110] km/s. However, as it was pointed out in *Lavie et al. (2017)*, this challenging for interpretation feature may be due to stellar variability.

Finally, the following physical picture, consistent also with the basic general estimates, follows from the undertaken numerical simulations. The supersonic interaction of SW with PW forms a bowshock ahead GJ436b. In the transition region between the ionopause and the bowshock the hydrogen atoms, carried in the PW stream, interact with the SW protons and generate ENAs. The absorption in Lyα observed for GJ436b is produced mostly by these ENAs, due to the resonant, or thermal, line broadening mechanism, augmented by the planetary hydrogen atoms, energized in the hot transition layer. The velocity of ENAs, like that of protons, is directed approximately away from the star. This results in a strong asymmetry of the absorption between the blue and the red wings of the Lyα line. It has to be noted, that no significant absorption comes from the regions beyond the compressed layer, because the hydrogen atoms are either too slow, or too rare there. Since the rate of ENAs production is proportional to the density of the planetary hydrogen atoms and stellar protons, the depth of the transit light-curve directly depends on the density of SW and intensity of the PW outflow. The PW intensity, in its turn, is proportional to the XUV flux. Because of the bowshock formation, the grazing transit due to orbital inclination is actually favorable to the absorption in Lyα, since the compressed region in this case transits across the center of stellar disk. The timing of early ingress is determined by the position of the bowshock ahead the planet and depends mostly on the SW density and PW intensity. At the same time, the shape of egress part of the transit light-curve is formed by the SW velocity.

The modeling, reported in this paper, quantitatively reproduces such observed features as the Lyα absorption line profiles at the mid-transit and post-transit phases (right panels in Figures 10 and 13, blue line), as well as the timing of early ingress and the maximum depth of the transit light-curve in the blue wing (Figures 14, 16, and 17). There is also an agreement between the simulation results and observations regarding the strong asymmetry of the absorption in the blue and red wings of the Lyα line and in the range of Doppler shifted velocities (<150 km/s), at which the absorption in the blue wing takes place. However, the model could not reproduce successfully the egress part of the transit light-curve. The suggested conclusion in that respect is twofold. It is possible that the available data are affected by strong variations of the SW parameters between the visits. At the same time, the effect of other physical factors, not yet included in the model, which may influence the interaction between the SW and PW around GJ436b, cannot be excluded as well. In particular, the magnetic fields of the stellar and planetary origin can be important. As demonstrated by the existing generic MHD simulations (e.g., *Bisikalo et al. 2013, Matsakos et al. 2015, Erkaev et al. 2017*), the presence of magnetic field extends the complexity of physics of the planetary environment and directly influences the interaction between PW and SW. It, therefore, brings an additional parameter for fitting of the modeling results to observations.

Altogether, the first application of the 3D hydrodynamic self-consistent multi-fluid model for the interpretation of transit observations of GJ436b, presented here, has shown that the concept of expanding and escaping upper atmospheres of hot exoplanets, developed earlier on the basis of 1D aeronomy models, gives adequate values for the density, velocity, and ionization degree of the outflowing PW material, and explains reasonably well the Lyα absorption measurements for GJ436b. The performed more realistic 3D simulations confirm the conclusion of previous 1D and 2D models, that the generation of ENAs during the interaction between the PW and SW plasmas is the main process responsible for the observed absorption in the Lyα line, whereas the radiation pressure effects play a secondary and less important role. It is also demonstrated that the Monte-Carlo modeling, used for the interpretation of observations in a number of cases before, has to be refined to include the parameters and geometry of the bowshock and/or ionopause, because the ENAs are generated in sufficient amount only in the compressed transition layer between them.

The remaining discrepancy between the simulated and observed Lyα transit light-curves and the line profiles indicates, on one hand, that the modeling, being basically correct and promising, admits further refining by taking into account of additional factors, on the other hand, that the model-based fitting of observations constrains the parameters of SW which, at the same time, may vary from one observational campaign to another.

**Appendix (1). Analytic estimations of the role of radiation pressure**

Close to the planet the escaping PW material is sufficiently dense to absorb efficiently the Lyα flux. If the self-shielding is strong, then it acts as a surface pressure in the same manner, as the SW ram pressure. The comparative role of both effects can be characterized by their ratio:

$$\frac{p_{Ly\alpha}}{p_{sw}} = \frac{F_{Ly\alpha}}{m_p n_{sw} V_{sw}^2} = \frac{L_{Ly\alpha}}{M'_{sw} V_{sw}} \quad , \tag{A1.1}$$

where $L_{Ly\alpha}$ is the total luminosity in the Lyα line and $M'_{sw}$, the total mass loss rate of star due to SW. Assuming typical values of $F_{Ly\alpha}$ =1 erg cm$^{-2}$ s$^{-1}$, at 1 a.u., $M'_{sw}$ ~2.5·10$^{11}$ g/s and $V_{sw}$ =10$^7$ cm/s, one obtains that the radiation pressure is about one order of magnitude less, than the SW ram pressure: $p_{Ly\alpha}/p_{SW}$ ~ 0.1

$P_{Ly\alpha}/P_{sw}$ ~ 0.1 .

Let's consider now the acceleration of neutrals by the radiation pressure only, i.e., disregarding the SW and gasdynamic interactions. The motion of a single atom in the stellar gravity field under the radiation pressure action can be easily solved, assuming the orbital momentum conservation. Equation for the radial velocity in the equatorial plane is:

$$m\frac{dV_R}{dt} = g(V)\beta m \frac{G}{R^2} - m\frac{G}{R^2} + m\frac{V_\varphi^2}{R} \tag{A1.2}$$

The azimuthal velocity can be expressed as $V_\varphi R = \text{const}$, $V_\varphi^2 = GR_{orb}/R^2$. Here we took into account that in the planet's frame of reference, from which the atom is launched, the centrifugal force is equal to gravity force. The factor $\beta$ in equation (A1.2) is the relation between the radiation pressure and gravity forces, g(V) is the stellar Lyα profile for which we take, for simplicity, but without loss of generality, the Gaussian profile, $g = \exp(-V^2/V_{Ly\alpha}^2)$. For GJ436, which radiates in Lyα line at the reference distance of 1 a.u. $F_{Ly\alpha}$ about 1 erg cm$^{-2}$ s$^{-1}$, the radiation force is a fraction of β=0.7 of the stellar gravity, while the line width is about 80 km/s (*Bourrier et al. 2015*). In the dimensionless form for GJ436b, with v=V$_R$/V$_{Ly\alpha}$, r=R/R$_{orb}$, where R$_{orb}$=4.35·10$^{11}$cm, and V$_{orb}$=120 km/s, the equation (A1.2) can be rewritten as follows:

$$\frac{dv^2}{dr} = \frac{\alpha}{r^2} \cdot \left[\beta \cdot e^{-v^2} - 1 + 1/r\right], \quad \alpha = 2V_{orb}^2/V_{Ly\alpha}^2 \approx 4.5 \qquad (A1.3)$$

Maximum terminal velocity, due to the finite line width, can be found by direct integration of equation $dv^2/dr = \beta \cdot e^{-v^2} \cdot \alpha/r^2$ with the result: $V_{max} = V_{Ly\alpha}\sqrt{\ln(1+2\beta V_{orb}^2/V_{Ly\alpha}^2)} \approx 95$ km/s. However, the actual acceleration of an atom by the radiation pressure force is restricted by the decrease of centrifugal force with distance, which compensates the gravity pull. An approximate analytic solution, which is sufficiently precise up to the maximum velocity, can be found by the series expansion of the $O(1-1/r)$ term in equation (A1.3):

$$v^2 \approx \ln\left[1 + \alpha\beta(1-1/r)\right] - 0.5\alpha(1-1/r)^2 + 0.25\alpha(1-1/r)^3 \qquad (A1.4)$$

Figure 18 shows the results of numerical integration of equation (A1.3) for the cases of β=0.7, 05, and 1. One can see that the maximum velocity does not exceed 60 km/s. It is reached rather far from the planet, at ~0.5R$_{orb}$~7R$_{star}$.

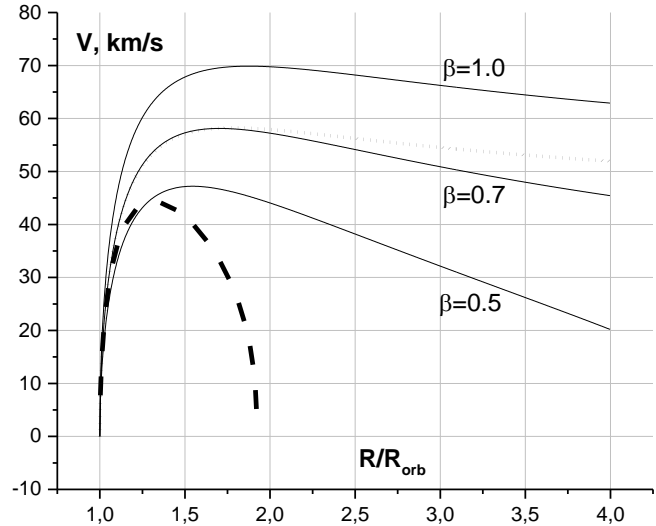

**Figure 18.** The radial velocity of a hydrogen atom as a function of distance, calculated by integration of equation (A1.3), for different values of the radiation pressure force. Dotted line shows the analytic solution (A1.4) for β=0.7. Dashed line shows the solution of an extended version of equation (A1.3) with the inclusion of ionization of hydrogen for the case of β=0.7.

Another factor which strongly decreases the impact of radiation pressure is the gasdynamic coupling between the neutrals and protons (*Shaikhislamov et al. 2016*). Due to charge exchange with a cross-section σ ~ 3·10⁻¹⁵ cm⁻², the hydrogen atoms effectively exchange momentum with protons over the distance of ~$R_p$ at the densities as low as n=10⁵ cm⁻³. Thus, when the density of protons exceeds that of atoms, the propulsion effect of the radiation pressure force on the plasma as a whole correspondingly decreases. This is especially true at small velocities. This effect can be taken into account by decreasing of the radiation pressure force (i.e., the parameter β) by a factor of $n_a/(n_a+n_p) \approx \exp(-t/\tau_{ion})$, which depends on the ionization time of hydrogen atoms. The time of particle's flight is found as $t = (R_{orb}/V_{Ly\alpha}) \int dr/v$. For the particular spectrum of GJ436b (*France et al. 2016*) with $\tau_{ion} = 29$ h, the dimensionless parameter $R_{orb}/(V_{Ly\alpha}\tau_{ion}) \approx 0.5$ is of the order of unity. The numerical solution of the extended version of equation (A1.3) with the inclusion of ionization is shown in Figure 18. One can see that photo-ionization reduces the velocity range of the accelerated by the radiation pressure hydrogen atoms down to values appeared within the interstellar (ISM) absorption and geocoronal contamination window. Therefore, these atoms will not contribute to the observed Lyα line absorption profile.

Altogether, the quantitative arguments presented above, show that the radiation pressure force of GJ436 is relatively insignificant, as compared to other forces, and alone it is not able to accelerate hydrogen atoms up to velocities necessary to produce the measured absorption in the Lyα line. This explains why in our simulations we see that the effect of the radiation pressure is very small and the main observational feature – the absorption of the Lyα line – is practically independent on it. To verify the performed estimates, we did also a simulation run with a very week SW ($M'_{sw}$ ~10¹⁰ g/s) and 20 times increased Lyα flux (~ 20 erg cm⁻² s⁻¹ at 1 a.u.). Only in this extreme case we saw the increased manifestation of the radiation pressure effect, resulting in a strong absorption with the depth comparable to that measured in observations.

**Appendix (2). Technical details of the numerical code of model**

The numerical scheme is explicit and uses up-wind donor cell method for flux calculations. To achieve second order spatial accuracy for differentials two grids are used shifted by half a step along each dimension. One grid is reserved for densities, temperatures, gravity potential and the other for velocities. For the second order temporal accuracy at each time step the code calculates at first (n,T) values using velocity field **V** and then calculates **V** using new (n,T) values. Numerical scheme fully conserves fluxes and total mass, and conserves Bernoulli constant along the characteristics. For energy simple non-conservative equation (3.3) is used. The energy conservation is checked by global integration and is used to evaluate the accuracy of simulation. Usually, the energy is balanced within 25%. We don't use any particular method to capture the shock. For the problems under consideration the accurate position of the shock and high front resolution are not crucial.

The spatial spherical grid uses uniform step for azimuth angle (in this study usually 96 points for a circumference, Δφ=0.065). The radial grid is exponential with step varying linearly with radius: $\Delta r = \Delta r_{min} + (\Delta r_{max} - \Delta r_{min}) \cdot (r - R_p)/(R_{max} - R_p)$. At the planet surface Δr is as small as at least $R_p$/200. For $\Delta r_{max}$ a value equal to $\Delta \varphi \cdot R_{max}$ is usually used. Thus, in the shock region ~20 $R_p$ the resolution is about $R_p$. For polar angle the grid is quadratic, $\theta = \Delta \theta_{min} \cdot i + \alpha \cdot i^2$, with smallest step at equatorial plane. Usually Δθ=Δφ at equator and 2Δφ at polar axis. Note that the exponential radial spacing in the spherical coordinate system allows keeping the same resolution in all 3 dimensions, if the azimuthal and latitudinal steps are chosen so that Δφ≈Δθ≈Δr/r. The influence of spatial resolution was checked by doubling the number of grid points for each dimension. The difference in results was sufficiently small.

For calculation of column densities along stellar rays we use numerically consuming but straightforward integration from the star to each cell in the planet spherical frame, using along the path the density values (but not the column densities itself) interpolated from nearby pixels. The calculated column density is used to determine attenuation of XUV flux in each spectral bin (0.1 nm). For Lyα the opacity is calculated in 30 km/s bins (in Doppler shifted units). Scattering of primary photons is not taken into account, while re-emitted photons are not considered because they do not produce a net radiation force anymore. To save numerical time, the radiation transfer is calculated usually each forth step of fluid dynamics and chemistry. We assume optically thin approximation for the photons generated by proton recombination to the ground state, so the total recombination coefficients are used.

The chemistry reactions are calculated by direct conversion of matrix $dn_i=R_{ji}(t,r) \cdot n_j \cdot n_i$ at each time step and at each pixel. This is not efficient numerically, but eliminates convergence problem due to widely different reaction rates $R_{ji}$.

The star boundary is treated in the same manner as the planet – by imposing the particular values. To launch the stellar wind more accurately, in a region around the star ($0.5R_{star}$) fixed values are prescribed according to the predetermined solution. Resolution around the star is rather coarse, a few points inside the star diameter, but it seems to be enough to simulate stellar wind.

For this particular study the runs continued for 500 dimensionless units which is about 60 orbits. As the main convergence criteria of steady state we use the total mass loss by the planet. After about 30 orbits it reaches asymptotic within 5%. We have not seen in simulations any disruption type variability due to interaction with stellar wind and instability of shock interface.

**Acknowledgements:**

This work was supported by grant № 18-12-00080 of the Russian Science Foundation. HL and MLK acknowledge the projects S11606-N16 and S11607-N16 of the Austrian Science Fund (FWF). MLK acknowledges the support from the FWF project I2939-N27. HL is also grateful to the FWF project P25256-N27. Parallel computing simulations, key for this study, have been performed at Computation Center of Novosibirsk State University, SB RAS Siberian Supercomputer Center, Joint Supercomputer Center of RAS and Supercomputing Center of the Lomonosov Moscow State University.

**References**

Arkhypov, O.V., Khodachenko, M.L., Lammer, H., et al., 2018, MNRAS, 476, 1224.

Ben-Jaffel, L. 2007, ApJL, 671, L61

Ben-Jaffel L., Sona Hosseini, 2010, ApJ, 709, 1284

Bisikalo D., Kaygorodov P., Ionov D., et al., 2013, ApJ, 764, 19.

Bisikalo, D. V., & Cherenkov, A. A., 2016, Astron. Reports, 60(2), 183-192.

Bourrier, V., Ehrenreich, D., Lecavelier des Etangs, A. 2015, A&A, 582, A65

Bourrier, V., des Etangs, A. L., Ehrenreich, D., et al., 2016, A&A, 591, A121.

Braginskii S. I., 1965, Rev. Plasma Phys., 1, 205

Cherenkov, A. A., Bisikalo, D. V., Kosovichev, A. G., 2017, MNRAS, 475(1), 605.

Christie, D., Arras, P., Li, Z. Y., 2016, ApJ, 820(1), 3

Daley-Yates, S., Stevens, I. R., 2018, MNRAS, 483(2), 2600.

Ehrenreich, D., Bourrier, V., Wheatley, P.J., et al., 2015, Nature, 522(7557), 459.

Erkaev, N. V., Odert, P., Lammer, H., et al., 2017, MNRAS, 470(4), 4330.

France K., Loyd R.P., Youngblood A., et al., 2016, ApJ, 820, 89

García Muñoz A. 2007, Planet. Space Sci., 55, 1426.

Holmström, M., Ekenbäck, A., Selsis, F., et al., 2008, Natur, 451, 970

Keppens, R., Goedbloed, J. P., 1999, arXiv preprint astro-ph/9901380.

Khodachenko M. L., Shaikhislamov I. F., Lammer H., et al., 2015, ApJ, 813, 50.

Khodachenko M. L., Shaikhislamov I. F., Lammer H., et al., 2017, ApJ, 847, 126.

Kislyakova, G. K., Holmström, M., Lammer, H., et al. 2014, Sci, 346, 981

Kislyakova, K. G., Holmström, M., Odert, P. et al., 2019, A&A, 623, A131

Koskinen T. T., Aylward A. D., Smith C. G. A., et al., 2007. ApJ, 661, 515

Koskinen T. T., Yelle R. V., Lavvas P., et al., 2010. ApJ, 723, 116

Koskinen, T. T., Harris, M. J., Yelle, R. V., et al., 2013, Icarus, 226(2), 1678-1694.

Kulow, J. R., France, K., Linsky, J., et al., 2014, ApJ, 786(2), 132.

Lammer H., Selsis F., Ribas I., et al., 2003, ApJ, 598, L121

Lavie B., Ehrenreich D., Bourrier V., et al., 2017, A&A, 605, L7

Lecavelier des Etangs, A., Vidal-Madjar, A., McConnell, J. C., et al. 2004, A&A, 418, L1

Lecavelier des Etangs, A., Vidal-Madjar, A., Desert, J.-M. 2008, Natur, 456, E1

Linsky, J. L., Yang, H., France, K., et al. 2010, ApJ, 717, 1291

Loyd R. P., Koskinen T.T., France K., et al., 2017, ApJ, 834, L17

Matsakos, T., Uribe, A., Königl, A., 2015, A&A, 578, A6.

Miller, S., Stallard, T., Tennyson, J., et al. 2013, JPCA, 117, 9770.

Owen, J. E., Adams, F. C., 2014, MNRAS, 444(4), 3761.

Shaikhislamov I. F., Khodachenko M. L., Sasunov Y. L., et al., 2014, ApJ, 795, 132

Shaikhislamov, I. F., Khodachenko, M. L., Lammer, H., et al., 2016, ApJ, 832, 173.

Shaikhislamov, I. F., Khodachenko, M. L., Lammer, H., et al., 2018a, ApJ, 866, 47.

Shaikhislamov, I.F., M.L. Khodachenko, H. Lammer, et al., 2018b, MNRAS, 481, 5315.

Tasitsiomi, A. 2006, ApJ, 645(2), 792

Trammell, G.B., Arras, P., Li, Z.-Y., 2011, ApJ, 728, 152.

Trammell G. B. Li Z.-Y. Arras P., 2014, ApJ, 788, 161.

Tremblin P. Chiang E., 2013, MNRAS, 428, 2565

Tripathi A., Kratter K. M., Murray-Clay R. A., et al., 2015, ApJ, 808, 173

Usmanov, A. V., William, H., Matthaeus, et al., 2011, ApJ, 727, 84

Vidal-Madjar A., Lecavelier des Etangs A., Désert J. M., et al., 2003, Nature, 422, 143

Vidal-Madjar, A., Désert, J., Lecavelier des Etangs, A., et al. 2004, ApJL, 604, L69

Yelle R. V., 2004, Icarus, 170, 167.